\documentclass[showpacs,amsmath,amsfonts,amssymb,aps,superscriptaddress]{revtex4}

\usepackage{tabularx}
\usepackage{enumitem}
\usepackage{graphicx}
\usepackage{float}
\usepackage{dcolumn}% Align table columns on decimal point
\usepackage{bm}% bold math
\usepackage{mathrsfs}

\usepackage{amsfonts}

\begin{document}

\title{Wheeler-DeWitt equation and the late gravitational collapse:
  effects of factor ordering and the tunneling scenario}
\author{Davide Batic}
\email{davide.batic@ku.ac.ae}
\affiliation{
Department of Mathematics,\\  Khalifa University of Science and Technology,\\ Main Campus, Abu Dhabi,\\ United Arab Emirates}
\author{M. Nowakowski}
\email{marek.nowakowski@ictp-saifr.org}
\affiliation{
ICTP South American Institute for Fundamental Research,\\ Rua Dr. Bento Teobaldo Ferraz 271, 
01140-070 S\~ao Paulo, SP Brazil }

\author{N. G. Kelkar}
\email{nkelkar@uniandes.edu.co}
\affiliation{Departamento de Fisica, Universidad de los Andes,\\
 Cra. 1E No. 18A-10, Bogota, Colombia}

\date{\today}

\begin{abstract}
We set up the Wheeler-DeWitt (WDW) equation for late gravitational
collapse. The fact that the gravitational collapse and the expanding/
collapsing universe can be described within the realm of the
Robertson-Walker metric renders the corresponding WDW equation for collapsing matter
a timeless Schr\"odinger equation. We explore the consequences of
such an equation and find the density to be quantized in terms of the
Planck density. Apart from that, the wave function as a solution of the
WDW equation shows that the initial singularity is avoided. We
concentrate on different factor orderings in the kinetic term of the
equation and show how after splitting off an exponential ansatz, new
polynomials entering the solution can be constructed. This enables us
to conclude that the factor ordering changes the details of the
solution and interpretation, but overall on a qualitative level the
results remain the same. We also probe into the effects of a positive
cosmological constant. It offers the possibility of a tunneling
scenario  at the cosmological horizon.
\end{abstract}
\maketitle
\section{Introduction}
The gravitational collapse is one of the most interesting phenomena which
General Relativity has to offer \cite {Weinberg, Joshi, Kent}. In its extreme case, 
it gives rise to compact objects like white dwarfs, neutron stars, and
black holes \cite{compactobjects}. At least in the case of black holes, 
we expect some effects and fingerprints of quantum gravity, as partly evidenced by
the Hawking radiation \cite{Hawking}.  Indeed, many suggestions have
been put forward in this direction \cite{Bambi, Malafarina}.  But one
obstacle in the line of research is the choice of the quantum gravity model.
Indeed, quantum gravity \cite{QuantumGravity} has already a history
close to a hundred years
\cite{Rocci, Rovelli} and yet we still cannot decide which one of the
many candidates \cite{Esposito} will lead to a consistent quantum
theory of gravitation.  Different points of view about the nature of
spacetime quantization and the lack of experimental data in this
region make quantum gravity an active area of research. One of the
persistent attempts was to consider General Relativity (GR) as a
field theory and quantize it canonically. The prerequisite for such an
undertaking is to identify dynamical variables \cite {ADM} and separate the
dynamics from constraints. The result of the canonical quantization is the 
WDW equation \cite{WDW} which has been used in quantum
gravity over fifty years \cite{WDWpapers1, WDWpapers2, WDWpapers3} and
remains popular until today.  The reason for this success is, on the
one hand, the fact that canonical quantization has proved otherwise to
be a powerful method in quantizing field theories and, on the other
hand, that the WDW equation for certain kind of metrics is relatively
simple to handle and can be partly interpreted. Regardless of the eventual form that a complete quantum gravity theory takes, it is probable that its connections to canonical quantization and the WDW equation will persist. Although not universally applied (see, for example, \cite{Chen1, Chen2}), the majority of studies exploring quantum gravity through the WDW equation utilize the Robertson-Walker (RW) metric. This approach is predominantly employed in the field of cosmology \cite{WDWcosmology}, where it serves to investigate questions concerning the origins of the universe. Clearly, one expects that quantum effects are
important in the early stages of the universe.
We emphasize here that the RW metric in GR is also
used in gravitational collapse \cite{Weinberg} which suggests that one
could examine also the WDW equation in the context of a gravitational
collapse.
Here, we would expect that quantum effects become significant as the collapse progresses towards smaller scale factors. In \cite{ourpaper}, we have studied such a scenario in its
simplest version, i.e., we chose the simplest scheme of the operator
factoring and put the cosmological constant to zero. The outcomes were notably promising: the initial singularity was circumvented, density became quantized, and the probability density revealed distinct sharp peaks. These peaks suggest a form of spacetime quantization on a probabilistic basis. Therefore, we revisited the topic to explore the effects of varying factor ordering, incorporate the cosmological constant, and examine a tunneling solution.

In cosmology, the WDW equation is frequently applied in scenarios devoid of matter, a condition that is untenable in the context of collapse. Consequently, we focus on identifying an appropriate Lagrangian, from which the classical Friedmann equations naturally emerge as Euler-Lagrange equations. This Lagrangian is
then used to perform the canonical quantization, culminating in the WDW equation
with matter.  Working under the assumption that all mass is behind the black hole horizon, we investigate the solutions to this equation across various factor ordering schemes. The most basic variant of this model has been examined in \cite{ourpaper}, yielding the promising outcomes described previously. Although it is not an eigenvalue equation, the requirement for square integrability leads to the quantization of density in terms of the Planck density. Given that factor ordering introduces an ambiguity in selection, investigating these various possibilities holds significant importance. In
the initial cases, we successfully solve the corresponding WDW
equation using an exponential ansatz. The resultant equation indeed
possesses polynomial solutions. This presents a quantization condition
analogous to the simplest case, and the physical interpretation
remains consistent. These polynomials allow us to construct the
complete wave function explicitly. As in the simplest case, the
initial singularity is avoided and the scale parameter has preferred
values in the probabilistic sense.

With the introduction of a positive cosmological constant, the potential within the WDW equation forms a barrier, setting the stage for an analysis of quantum tunneling. We estimate the tunneling probability, interpreting it as the likelihood of tunneling at the cosmological horizon. Given that in the presence of a positive cosmological constant, the Hamiltonian becomes unbounded, assessing the self-adjointness of this operator is crucial. This assessment is conducted in a dedicated subsection

We draw the reader's attention to the fact that several other papers
examined the gravitational collapse in the context of a
Schro\"{o}dinger-like equation. In \cite{Rosen, Feleppa}, the divergence from our methodology lies in the Hamiltonian, which distinctly is not the WDW operator. Furthermore, we highlight the notable absence of a time variable in the WDW equation, a prevalent issue in quantum gravity also encountered in the canonical quantization framework \cite{time1, time2, time3, time4, time5, time6, time7, time8, time9, time10, time11, time12}. Contrarily, in studies such as \cite{Rosen, Feleppa}, and others like \cite{Stojkovic1, Stojkovic2, Pal}, Schr\"{o}dinger-like equations in the form $i\partial_t \Psi= H\Psi$ are presented, which starkly contrast with the WDW formulation $H\Psi=0$. Thus, these methodologies diverge from ours right from the outset.

The paper is organized as follows. In sections II and III, we list the
    salient features of \cite{ourpaper} where the simplest scheme of
    factor ordering was used. In section IV,  we study a more general
    factor ordering and compare the outcome with our previous
    results. Section V is devoted to effects when the cosmological
    constant $\Lambda$ is included. We analyze the resulting potential and the self-adjointness of the
  WDW operator. Next we give a general quantization condition for
  the density in the presence of this constant. We outline a tunneling
scenario connected with $\Lambda$. In section VI, we give our conclusions.

\section{The WDW equation in the gravitational collapse}

As we mentioned in the introduction, the gravitational aspects of
collapsing matter and the salient details of a universe are described
by the Friedmann-Robertson-Walker line element with $c=1$ \cite{Weinberg}
\begin{equation}\label{LE}
ds^2=dt^2-R^2(t)\left[\frac{dr^2}{1-kr^2}+r^2 d\vartheta^2+r^2\sin^2{\vartheta}d\varphi^2\right].    
\end{equation}
However, in cosmology, the curvature parameter $k$  takes on values of
$\pm 1,0$, rendering $r$ as dimensionless. Meanwhile, the scale factor
$R$ carries a dimension of length.  In the collapse scenario for cold dust, we have \cite{Weinberg}
\begin{equation}\label{kappa}
k=\frac{8\pi}{3}G\rho_0,    
\end{equation} 
where $G$ is the Newton's gravitational constant and $\rho_0$ is the
density of a spherically symmetric, isotropic and homogeneous
matter distribution. Considering that 
our main objective is to explore quantum effects in the latter stages of collapse, we note the improbability of densities conforming to the conventional values typically associated with the onset of collapse, as documented in \cite{Bode}. Our emphasis is, therefore, on the post-black hole formation phase of the collapse. In natural units, where $\hbar=c=1$, it is obvious that $k$ has the dimension $L^{-2}$. The apparent mismatch in dimensions between the cosmological scale factor $R$ and the same factor in the context of gravitational collapse can be easily remedied by the following rescaling 
\begin{equation}\label{rescaling}
a(t)=L_0 R(t),\quad
\widetilde{r}=\frac{r}{L_0},\quad L_0=\frac{1}{\sqrt{k}}.
\end{equation}
As $L_0$ carries the dimension of length, the scale factor $a$ now possesses the same dimension. This allows us to recast the line element (\ref{LE}) as
\begin{equation}\label{LE1}
ds^2=dt^2-a^2(t)\left[\frac{d\widetilde{r}^2}{1-\widetilde{r}^2}+\widetilde{r}^2 d\vartheta^2+\widetilde{r}^2\sin^2{\vartheta}d\varphi^2\right].
\end{equation}
This transformation makes possible to draw parallels between gravitational
collapse and cosmology when $k=1$ in the latter case. Especially, the
derivation of the WDW equation without matter will be the same in both
cases. For this purpose, we follow the methodology outlined in \cite{KT,Inverno}.  We refer for details to \cite{ourpaper} and quote here the final result in terms of
the action
\begin{eqnarray}
S&=&\frac{3\pi}{4G}a^2\dot{a}-\frac{3\pi}{4G}\int dt\left[a\dot{a}^2-a+\frac{\Lambda}{3}a^3\right]-\frac{3\pi}{4G}a^2\dot{a}+S_{matter},\\
&=&-\frac{3\pi}{4G}\int dt\left[a\dot{a}^2-a+\frac{\Lambda}{3}a^3\right]+S_{matter}.
\end{eqnarray}
We highlight here that constructing a matter Lagrangian that yields
the energy-momentum tensor of a perfect fluid is not a straightforward
enterprise \cite{Schutz,Ray}. However, if we carefully constrain the
variation $\delta g_{\mu\nu}$, we find that a suitable candidate is provided by
\begin{equation}
    S_{matter}=\int_{\mathcal{M}}d^4 x\sqrt{-g}L_{matter}=-\int_{\mathcal{M}}d^4 x\sqrt{-g}\rho=-2\pi^2\int a^3\rho dt.
\end{equation}
It results in a total Lagrangian expressed by the equation
\begin{equation}
L=-\frac{3\pi}{4G}\left(a\dot{a}^2-a+\frac{\Lambda}{3}a^3\right)-2\pi^2a^3\rho.
\end{equation}
Indeed, one can show \cite{ourpaper} that this Lagrangian leads to the
correct Friedmann equation and is suitable for quantization. Taking
into account that the Hamiltonian $H=\pi_a\dot{a}-L$ has conjugate
momentum \cite{KT} (we take the opportunity to point out that there is a
typographical error in \cite{ourpaper} in connection with this equation)
\begin{equation}
\pi_a=-\frac{3\pi}{2G}a\dot{a},    
\end{equation}
it is straightforward to derive the Hamiltonian
\begin{equation}\label{Hamiltonian}
H=-\frac{G}{3\pi}\frac{\pi_a^2}{a}+\frac{3\pi}{4G}\left(-a+\frac{\Lambda}{3}a^3\right)+2\pi^2 a^3\rho.
\end{equation}
Applying the canonical quantization prescription $\pi_a\to -id/da$, the  WDW equation $H\Psi(a)=0$ becomes
\begin{equation}\label{SCH1}
\left[\frac{G}{3\pi a}\frac{d^2}{da^2}+\frac{3\pi}{4G}\left(-a+\frac{\Lambda}{3}a^3\right)+2\pi^2 a^3\rho\right]\Psi(a)=0. 
\end{equation}
For later purposes, it is of some importance to note that we have chosen a factor ordering 
\begin{equation}\label{fa}
\pi^2_a\to-\frac{d^2}{da^2}.    
\end{equation}
The consequences of different factor ordering
choices in the context of gravitational collapse will be further
explored in the following section. Finally, we can cast (\ref{SCH1})
in the form
\begin{equation}\label{WDW01}
\left(-\frac{d^2}{da^2}+V_{eff}(a)\right)\Psi(a)=0,\quad
V_{eff}(a)=-\frac{9\pi^2}{4G^2}\left(-a^2+\frac{\Lambda}{3}a^4\right)-\frac{6\pi^3}{G}a^4\rho.
\end{equation}
An examination of dimensions reveals that all terms within the round
brackets of equation (\ref{WDW01}) share a consistent dimension of
$M^2$. This is in contrast to \cite{Braz} where a similar equation has been independently derived through different means. In the context of cold dust, where $\rho=\rho_0 (a_0/a)^3$, the corresponding WDW equation is 
\begin{equation}
\left[-\frac{d^2}{da^2}+\left(-\frac{3\pi^2\Lambda}{4G^2}a^4+\frac{9\pi^2}{4G^2}a^2-\frac{6\pi^3\rho_0 a_0^3}{G}a\right)\right]\Psi(a)=0.
\end{equation}
Coming back to the original variables $R\equiv \widetilde{a}=a/L_0$
with $L_0$ defined as in (\ref{rescaling}), the above equation takes
the form
\begin{equation}\label{WDWp}
\left[-\frac{d^2}{d\widetilde{a}^2}+\left(-\frac{3\pi^2\Lambda L_0^6}{4G^2}\widetilde{a}^4+\frac{9\pi^2 L_0^4}{4G^2}\widetilde{a}^2-\frac{6\pi^3\rho_0 \widetilde{a}_0^3 L_0^6}{G}\widetilde{a}\right)\right]\Psi(\widetilde{a})=0.
\end{equation}
Furthermore, if we normalize the radial coordinate $r$ so that
$\widetilde{a}_0=1$ \cite{Weinberg} and introduce the Planck density $\rho_{Pl}=1/G^2$
as well as the vacuum density associated with the cosmological constant $\rho_{vac}=\Lambda/(8\pi G)$ then (\ref{WDWp}) becomes
\begin{equation}\label{WDWfin}
\left(-\frac{d^2}{d\widetilde{a}^2}+U_{eff}(\widetilde{a})\right)\Psi(\widetilde{a})=0
\end{equation}
with 
\begin{equation}\label{Veff}
U_{eff}(\widetilde{a})=\alpha \widetilde{a}^4+\beta\widetilde{a}(\widetilde{a}-1).
\end{equation}
The parameters are defined as follows
\begin{equation}\label{coeff}
\alpha=-\frac{\rho_{vac}}{\rho_0}\beta,\quad
\beta=\frac{81}{256}\left(\frac{\rho_{Pl}}{\rho_0}\right)^2.
\end{equation}

\section{Solution of the WDW equation in the case $p=0$}
In our analysis, we neglect the cosmological constant \(\Lambda\) due to its comparatively negligible contribution to the effective potential. To clarify, the parameter \(\beta\) defined in (\ref{coeff}) is extremely large due to the significant difference between the Planck density (\(\rho_{\text{Pl}} \sim 5.1 \times 10^{96} \, \text{kg/m}^3\)) and the initial density of the collapsing dust cloud (\(\rho_0 \sim 10^{-16} \, \text{kg/m}^3\)). Conversely, the vacuum energy density associated with \(\Lambda\) (\(\rho_{\text{vac}} \sim 5.9 \times 10^{-27} \, \text{kg/m}^3\)) is many orders of magnitude smaller. Therefore, the term involving \(\Lambda\) in the effective potential is insignificant compared to the terms involving \(\beta\), justifying its omission in this context. The simplified effective potential is thus given by
\begin{equation}\label{VeffL}
U_{eff}(\widetilde{a})=\beta \widetilde{a}(\widetilde{a}-1),   
\end{equation}
which represents a parabola with a minimum at $\widetilde{a}=1/2$. The
equation at hand is not an eigenvalue equation for a shifted harmonic
oscillator, but it is clear that there exists a connection. Indeed, 
by introducing the transformation $\widehat{a}=\widetilde{a}-1/2$, we
can recast the WDW equation into
\begin{equation}\label{WDW0}
  -\frac{d^2\Psi}{d\widehat{a}^2}+\beta\widehat{a}^2\Psi(\widehat{a})=\frac{\beta}{4}\Psi(\widehat{a}).
\end{equation}
subject to the boundary condition $\Psi(\widetilde{a})\to 0$ as $\widetilde{a}\to{+\infty}$  and with the normalization condition
$\int_0^\infty|\Psi(\widetilde{a})|^2~d\widetilde{a}=1$. These constraints are motivated by the requirement of square integrability and the physical interpretation of the wave function. The condition at infinity ensures that the wave function is normalizable, reflecting the probabilistic nature of quantum mechanics, while the condition at $a=0$ corresponds to avoiding the classical singularity, in line with the expectation that quantum gravity effects prevent such singularities. Furthermore, these boundary conditions are analogous to those used in the quantum harmonic oscillator problem, where normalizable solutions must vanish at infinity.

In this reformulation, $\beta$ emerges as a characteristic parameter, thereby playing a fundamental role in the governing wave equation. If we introduce the dimensionless variable $\xi=\sqrt[4]{\beta}\widehat{a}$, (\ref{WDW0}) becomes 
\begin{equation}\label{Herm}
\frac{d^2\Psi}{d\xi^2}=\left(\xi^2-K\right)\Psi(\xi),\quad
K=\frac{\sqrt{\beta}}{4}.
\end{equation}
This form is reminiscent of a dimensionless harmonic oscillator
\cite{Grif}. This leads to the following ansatz for the unnormalized wave function
\begin{equation}
\Psi(\xi)=h(\xi)e^{-\xi^2/2},    
\end{equation}
which applied to (\ref{Herm}) leads to the Hermite differential equation
\begin{equation}
\frac{d^2 h}{d\xi^2}-2\xi\frac{dh}{d\xi}+(K-1)h(\xi)=0.    
\end{equation}
Since the requirement for square integrability is satisfied when $K=2n+1$, it leads to
\begin{equation}\label{quant0}
\beta=16(2n+1)^2,\quad n=0,1,2,\cdots.
\end{equation}
In other words we obtain a quantization for the density in the form
\begin{equation}\label{q0}
\rho_0\equiv \rho_{0, n}=\frac{9\rho_{Pl}}{64(2n+1)},\quad n=0,1,2,\cdots.
\end{equation}
Finally, it can be easily checked that the ground state wave function is 
\begin{equation}\label{zeroeig}
\psi_0(\widetilde{a})= c_0 e^{-2\left(\widetilde{a}-\frac{1}{2}\right)^2},\quad c_0=\frac{\sqrt{2}}{\sqrt{\sqrt{\pi}\left[1+\mbox{erf}(1)\right]}}.
\end{equation}
\begin{figure}[ht!]
    \includegraphics[width=0.3\textwidth]{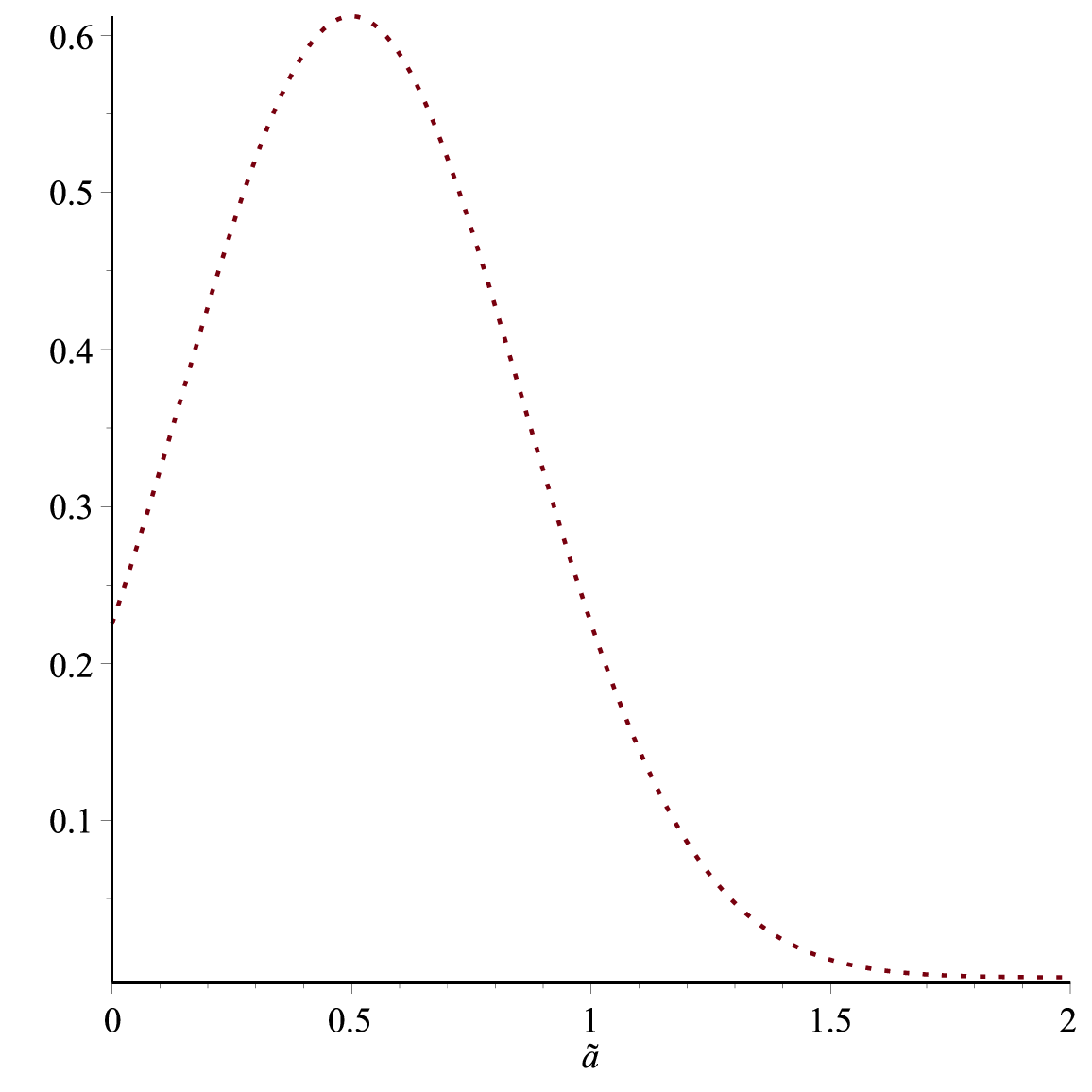}
    \includegraphics[width=0.3\textwidth]{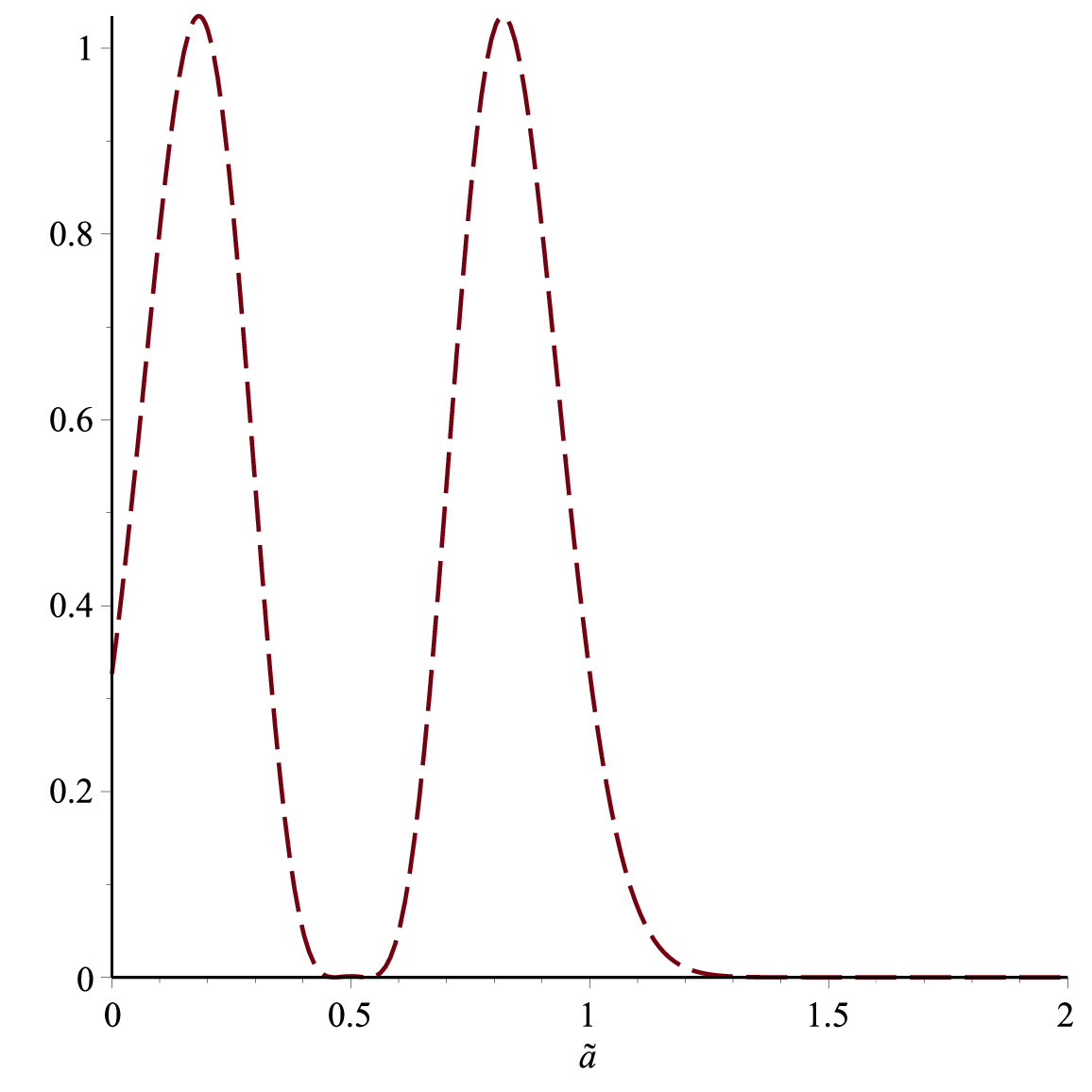}
    \includegraphics[width=0.3\textwidth]{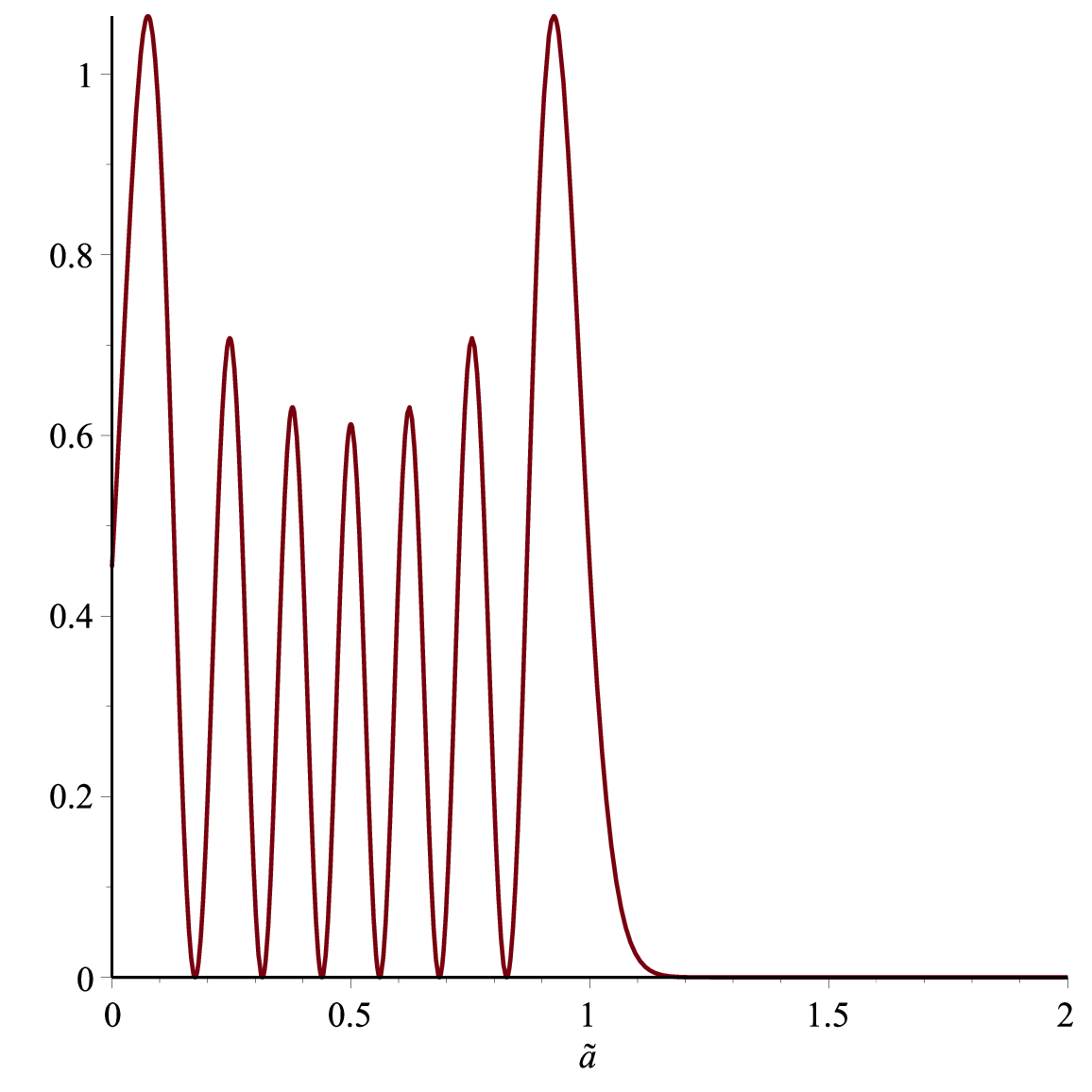}
\caption{\label{figure00}
Plot of the probability density $|\Psi_n|^2$ for the ground state $n=0$ (dotted line), $n=2$ (dashed line) and $n=6$ (solid line).}
\end{figure}

As $n$ increases, the discrepancy between successive $\rho_{0,n}$
values diminishes, eventually approaching a continuum, a trend
supported by the probability density $|\Psi_n|^2$ as depicted in
Figure~\ref{figure00}. This indicates that at lower $n$ values, the
scale parameter exhibits preferred positions in probabilistic terms,
while at higher $n$, it transitions back to a continuous
spectrum. Like many phenomena related to quantum effects in the realm
of black hole physics, empirical verification remains
elusive. However, we expect that quantum gravity theories will
naturally avoid the central singularity, an outcome that is in line with
various theoretical frameworks, including the Fuzzball paradigm in string theory \cite{Mathur1,Mathur2,Bena}, which states that singularities are replaced by complex quantum states (fuzzballs). Although our approach does not explicitly adopt the Fuzzball paradigm, the avoidance of singularities through quantum gravitational effects is a common theme \cite{Struyve2017,Bojowald2007,Singh2024}. Both our results and the Fuzzball paradigm suggest that a complete theory of quantum gravity will fundamentally change our understanding of black hole interiors.

A speculative note concerns the presence of the wave function beyond the value $1$: if all matter is enclosed within the black hole horizon, the value $1$ denotes the
horizon boundary \cite{ourpaper}. This can be seen by starting
  with $\rho_0+M_0/(4\pi/3)a_s^3$. Using the latter and writing
  $a_s=L_o\tilde{a}_s$, one can infer that $\tilde{a}_s=1$ corresponds
  to $a_s=2GM_0$.
Thus, the wave function's extension beyond this value could be associated with phenomena such as Hawking radiation \cite{ourpaper}. Alternatively, it may indicate that the black hole is transitioning to a ground state characterized by $n=0$, potentially resulting in a black hole remnant.

In conclusion, we end this section by noting a typographical error in equation (43) as cited in \cite{ourpaper}. The argument of the Hermite polynomials presented therein should be written as $2\sqrt{2n+1}(\widetilde{a}-1/2)$.

\section{Effects of different factor orderings}
Our focus now shifts to examining how the results of the previous section are influenced by variations in the choice of factor ordering. The ambiguity in factor ordering is a critical aspect to explore, as it is essential to determine whether it leads to qualitative consistency or discrepancy in our findings. The process of quantization inherently involves replacing momenta with differential operators, introducing uncertainties. This necessitates the inclusion of a factor ordering parameter $p$ in the WDW equation, following the prescription outlined in the literature \cite{KT,Hartle,Vilenkin}
\begin{equation}\label{factorord}
 \pi^2_a \rightarrow -a^{-p}\left[\frac{d}{da}a^p\frac{d}{da}\right] = -\left(\frac{d^2}{da^2} + \frac{p}{a}\frac{d}{da}\right).
\end{equation}
For the purpose of achieving congruence with empirical observations, it is essential to meticulously address the intrinsic ambiguity in the WDW equation. This involves discerning the operator ordering factor through the imposition of relevant constraints. A noteworthy candidate for the ordering factor is $p=1$, rendering (\ref{factorord}) similar to the radial component of the Laplacian operator \cite{Page,Kont}. Furthermore, \cite{Vil}  derived an exact solution for the WDW equation when subjected to this particular choice. Alternatively, \cite{Gao}, under the premise that the universe wave function is finite, advocated for the adoption of $p=-2$. Additionally, \cite{Vieira} postulated a boundary condition which stipulates that the ordering parameter be confined within the range $0\leq p\leq 2$.  It is imperative to emphasize, as pointed out by \cite{Vil,He1,He2}, that the effect of operator ordering becomes prominent only when the universe is relatively small, indicating that the factor ordering issue cannot be resolved purely through the semiclassical limit \cite{Steigl}. Moreover, it is relevant to acknowledge the observations by \cite{Vil}, which cast doubt on the conclusiveness of the arguments presented by  \cite{Page} in favor of a particular choice for $p$. Consequently, at this step, we will exercise restraint in committing to a specific value of $p$. A definitive choice will be contingent upon the solution of the WDW equation satisfying the boundary conditions introduced immediately after equation (\ref{WDW0}).

\begin{figure}[!ht]
    \includegraphics[width=0.3\textwidth]{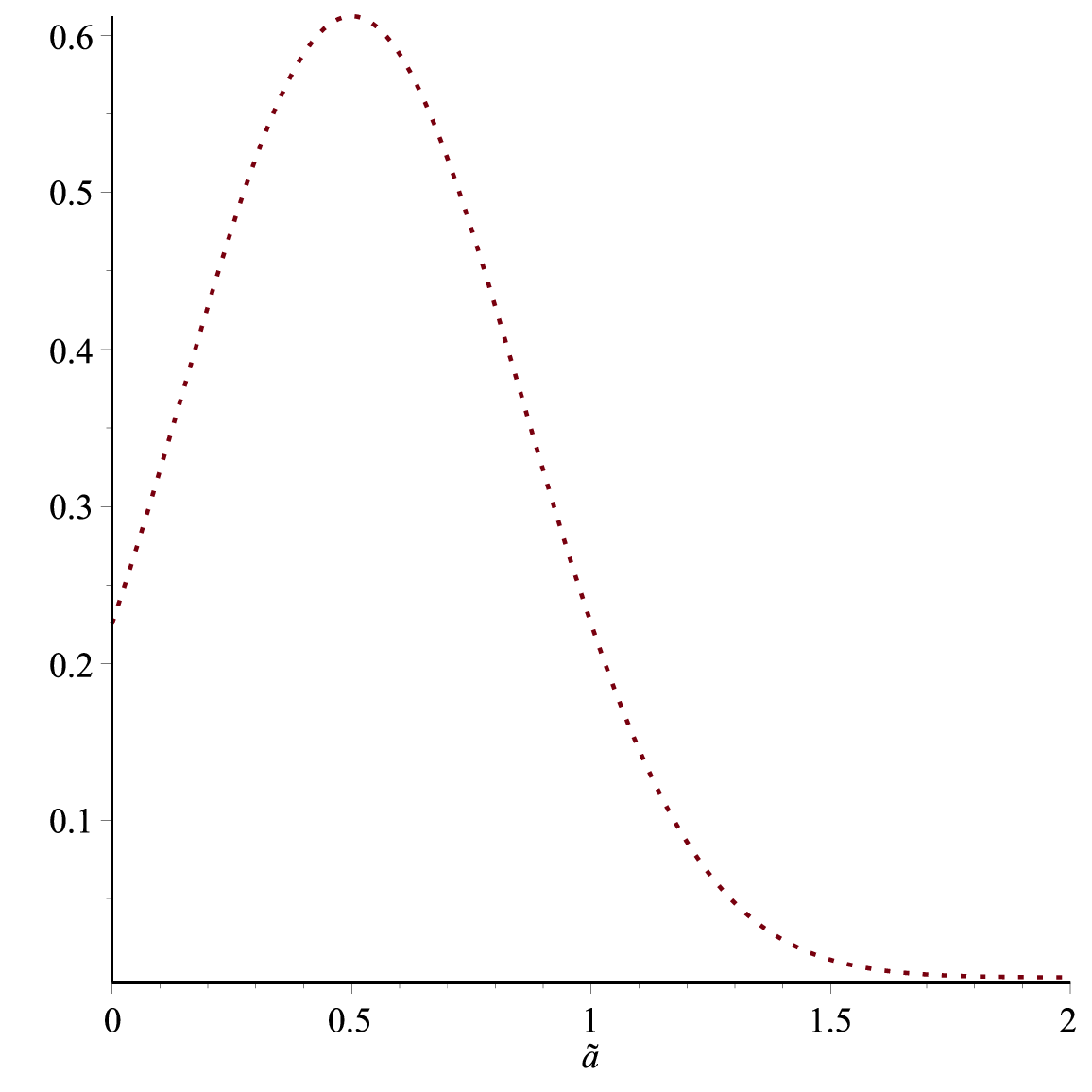}
    \includegraphics[width=0.3\textwidth]{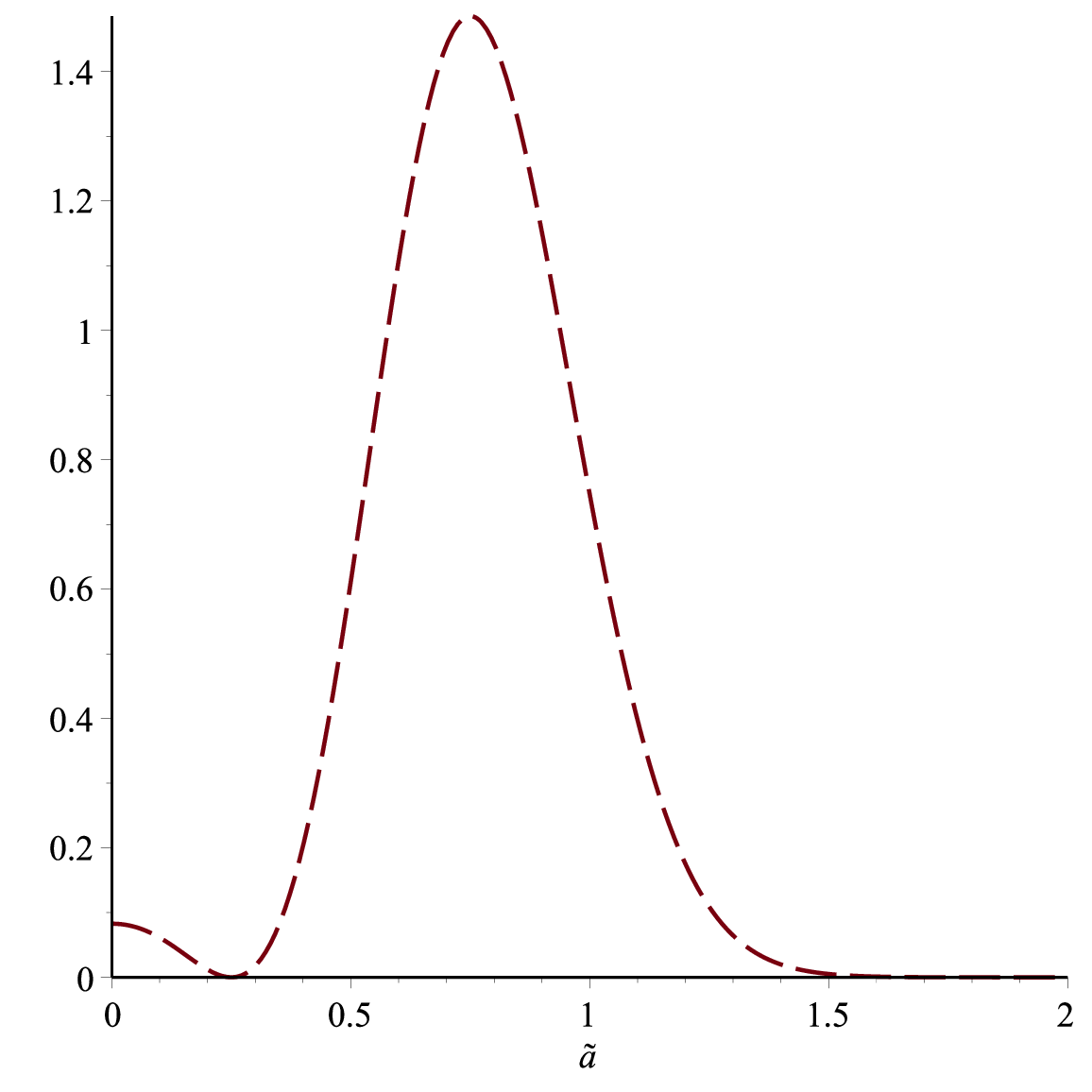}
    \includegraphics[width=0.3\textwidth]{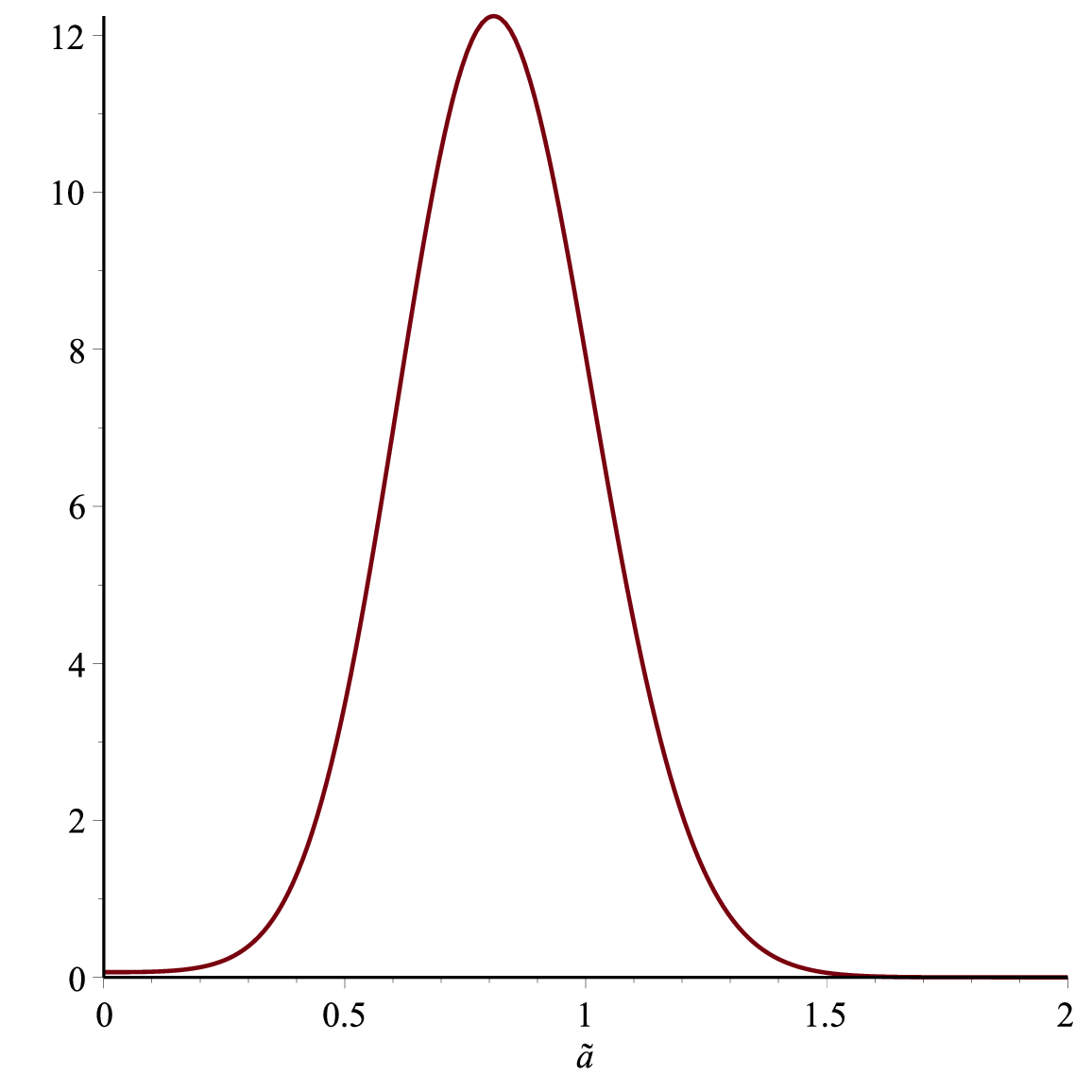}
\caption{\label{figure1geon}
Plots of the probability density $|\Psi_{p_i}|^2$ for $p_0=0$ (dotted line), $p_1=-1$ (dashed line), $p_2=-5+\sqrt{5}$ (solid line).}
\end{figure}

\begin{figure}[!ht]
    \includegraphics[width=0.3\textwidth]{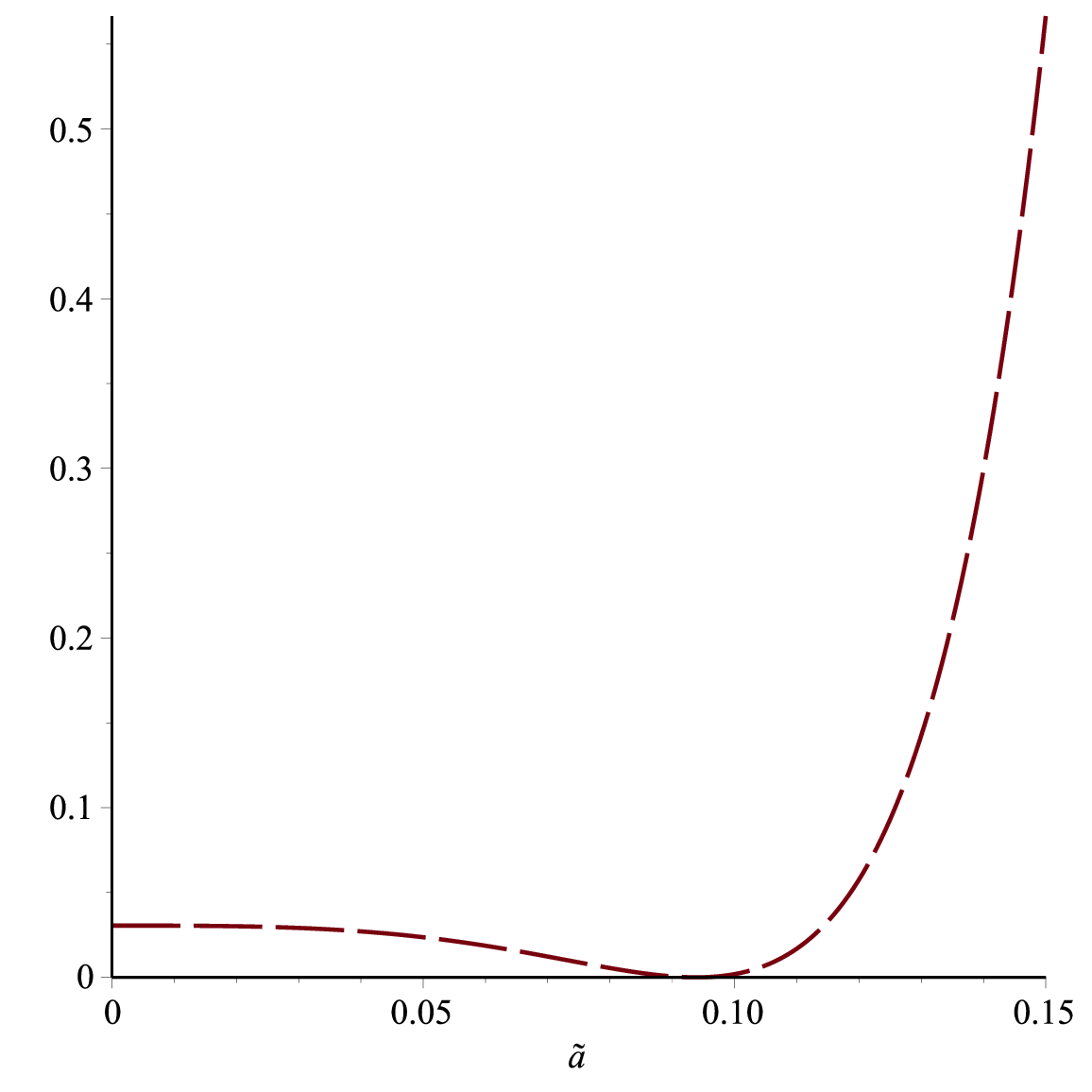}
    \includegraphics[width=0.3\textwidth]{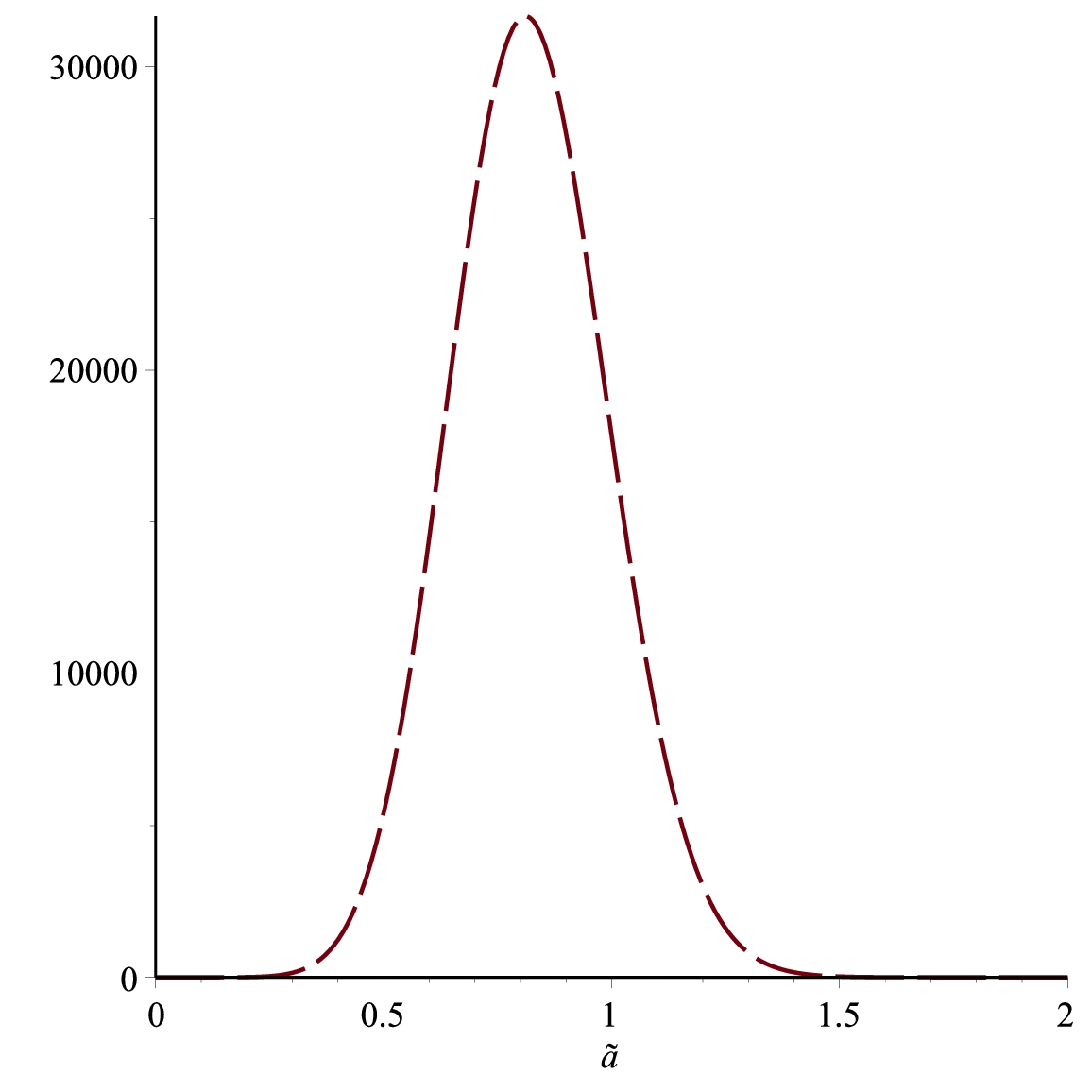}
    \includegraphics[width=0.3\textwidth]{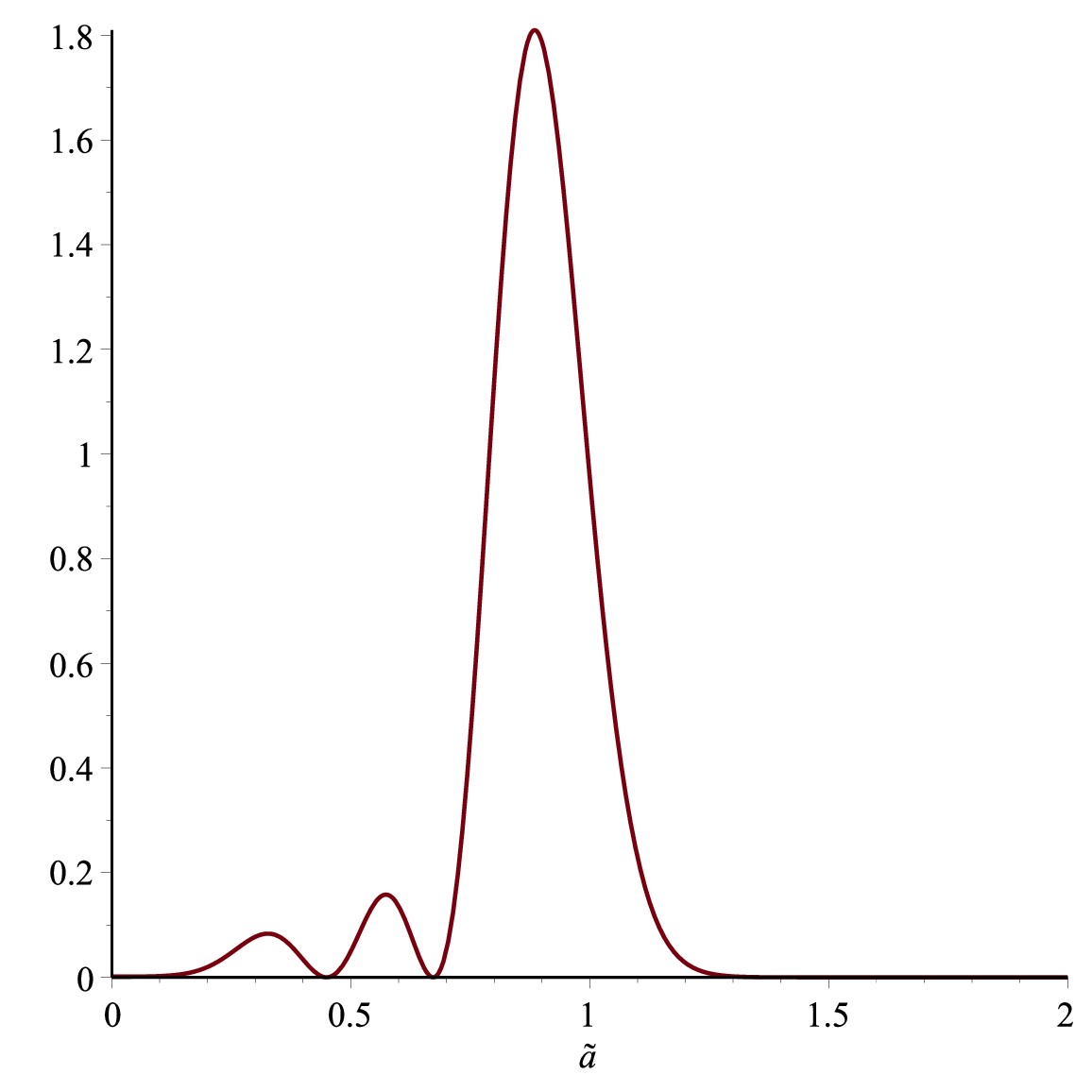}
\caption{\label{figure1geon1}
Plots of the probability density $|\Psi_{p_i}|^2$ for $p_0 =-4$ (long dashed line) and $p_4=--2.86171$ (solid line).}
\end{figure}

For the generalized prescription as illustrated in (\ref{factorord}), we can derive the corresponding WDW equation for cold dust from (\ref{fa}). The result is as follows
\begin{equation}
\left[\frac{d^2}{da^2}+\frac{p}{a}\frac{d}{da}+\left(\frac{3\pi^2\Lambda}{4G^2}a^4-\frac{9\pi^2}{4G^2}a^2+\frac{6\pi^3\rho_0 a_0^3}{G}a\right)\right]\Psi(a)=0.
\end{equation}
Adhering to the approach adopted in the previous section, we apply the rescaling $\widetilde{a}=a/L_0$, in conjunction with the Planck and vacuum densities. This allows us to rewrite the above equation as
\begin{equation}\label{equaz}
\frac{d^2\Psi}{d\widetilde{a}^2}+\frac{p}{\widetilde{a}}\frac{d\Psi}{d\widetilde{a}}-\left[\alpha\widetilde{a}^4+\beta\widetilde{a}(\widetilde{a}-1)\right]\Psi(\widetilde{a})=0.
\end{equation}
Here, $\alpha$ and $\beta$ are defined as in (\ref{coeff}). For the remaining part of this section, we are particularly interested in the case when $\Lambda=0$, hence, we will set $\alpha=0$. In order to proceed with our analysis, we start with the following ansatz
\begin{equation}\label{tr}
\Psi(\widetilde{a})=e^{-\frac{\sqrt{\beta}}{2}\left(\widetilde{a}-\frac{1}{2}\right)^2}g(\widetilde{a}).
\end{equation}
In this context, we focus on the cases where the unknown function $g(\widetilde{a})$ is a polynomial. We can see that the combination of equations (\ref{equaz}) and (\ref{tr}) leads to the following differential equation
\begin{equation}
\widetilde{a}\frac{d^2 g}{d\widetilde{a}^2}+(p+\sqrt{\beta}\widetilde{a}-2\sqrt{\beta}\widetilde{a}^2)\frac{dg}{d\widetilde{a}}+\left[\frac{p}{2}\sqrt{\beta}+\left(\frac{\beta}{4}-\sqrt{\beta}-p\sqrt{\beta}\right)\widetilde{a}\right]g(\widetilde{a})=0.
\end{equation}
We initially observe that the constant polynomial solution $g(a)=g_0\neq 0$ is attained for $\sqrt{\beta}=4(p+1)$ and $p=p_0=0$. These conditions correspondingly yield $\beta=16$. Encouragingly, as anticipated, this solution indeed reproduces the eigenfunction given by (\ref{zeroeig}) with $g_0=c_0$. Building on this initial result, we now extend our investigation to encompass additional solutions, such as $g(\widetilde{a})=g_0+g_1\widetilde{a}$. In this case, we find that the latter is indeed a solution, provided that
\begin{equation}
\sqrt{\beta}=4(p+3),\quad 
g_1=-\frac{\sqrt{\beta}}{2}g_0=-2(p+3)g_0,\quad 
p^2+5p+4=0.
\end{equation}
The latter equation has roots $p=-1$ and $p=-4$. The last root must be disregarded in view of the fact that the condition $\sqrt{\beta}>0$ requires that $p>-3$. Hence, $p=p_1=-1$ and the corresponding eigenfunction is
\begin{equation}\label{sol2}
\Psi_{p_1}(\widetilde{a})=g_0e^{-4\left(\widetilde{a}-\frac{1}{2}\right)^2}(1-4\widetilde{a}).  
\end{equation}
After the examination of the case of first-degree polynomials, we will now tackle the construction of quadratic polynomials of the form $g(\widetilde{a})=g_0+g_1\widetilde{a}+g_2\widetilde{a}^2$. We find that
\begin{equation}
\sqrt{\beta}=4(p+5),\quad 
g_1=-\frac{\sqrt{\beta}}{2}g_0=-2(p+5)g_0,\quad 
g_2=\frac{\sqrt{\beta}}{8}\left(4+\sqrt{\beta}\right)g_0=2(p+5)(p+6)g_0,\quad
(p+5)(p^2+10p+20)=0.
\end{equation}
Since $p>-5$, the only acceptable root is $p=p_2=-5+\sqrt{5}$. The corresponding eigenfunction is given as follows
\begin{equation}\label{sol3}
\Psi_{p_2}(\widetilde{a})=g_0e^{-\sqrt{20}\left(\widetilde{a}-\frac{1}{2}\right)^2}\left[1-2\sqrt{5}\widetilde{a}+2\sqrt{5}(1+\sqrt{5})\widetilde{a}^2\right].
\end{equation}
The same method applied thus far allows for the construction of additional polynomial solutions of increasingly higher degrees. In what follows, we limit us to summarize the corresponding results in the case of cubic and quartic polynomial solutions
\begin{enumerate}
\item 
    $g(\widetilde{a})=\sum_{n=0}^3 g_n\widetilde{a}^n$: we have $\sqrt{\beta}=4(p+7)$ and
    \begin{eqnarray}
     g_1&=&-\frac{\sqrt{\beta}}{2}g_0=-2(p+7)g_0,\\
     g_2&=&\frac{\sqrt{\beta}}{8}\left(4+\sqrt{\beta}\right)g_0=2(p+7)(p+8)g_0,\\
     g_3&=&-\frac{\beta^{3/2}(p+2)+12p\beta+40\beta}{48(p+2)}g_0=
     -\frac{4(p+7)^2(p^2+12p+24)}{3(p+2)}g_0.
    \end{eqnarray}
    The only root of the equation
    \begin{equation}
    p^5+32p^4+391p^3+2268p^2+6204p+6384=0    
    \end{equation}
    satisfying the condition $p>-7$ is $p=p_3=-4$. The corresponding eigenfunction reads
    \begin{equation}\label{sol4}
    \Psi_{p_3}(\widetilde{a})=g_0e^{-6\left(\widetilde{a}-\frac{1}{2}\right)^2}\left(1-6\widetilde{a}+24\widetilde{a}^2-784\widetilde{a}^3\right].
    \end{equation}
\item
    $g(\widetilde{a})=\sum_{n=0}^4 g_n\widetilde{a}^n$: we have $\sqrt{\beta}=4(p+9)$ and
    \begin{eqnarray}
     g_1&=&-\frac{\sqrt{\beta}}{2}g_0=-2(p+9)g_0,\\
     g_2&=&\frac{\sqrt{\beta}}{8}\left(4+\sqrt{\beta}\right)g_0=2(p+9)(p+10)g_0,\\
     g_3&=&-\frac{\beta^{3/2}(p+2)+12p\beta+40\beta}{48(p+2)}g_0=
     -\frac{16(p+9)(p+10)}{p+6}g_0,\\
     g_4&=&\frac{2(p+9)^2(p^4+29p^3+280p^2+1032p+1272)}{3(p^2+5p+6)}g_0.
    \end{eqnarray}
    The only root of the equation
    \begin{equation}\label{sol5}
    p^7+52p^6+1121p^5+12950p^4+86160p^3+328344p^2+661776p+543744=0    
    \end{equation}
    satisfying the condition $p>-9$ is $p=p_4=-2.86171$. The corresponding eigenfunction is
    \begin{equation}\label{psip4}
    \Psi_{p_4}(\widetilde{a})=g_0e^{-12.276\left(\widetilde{a}-\frac{1}{2}\right)^2}\left(1-12.276\widetilde{a}+87.634\widetilde{a}^2-223.392\widetilde{a}^3+173.903\widetilde{a}^4\right].
    \end{equation}
\end{enumerate}
All constructed solutions are regular at $\widetilde{a}=0$ and they satisfy the boundary condition $\Psi(\widetilde{a})\to 0$ as $\widetilde{a}\to{+\infty}$ with the normalization condition $\int_0^\infty|\Psi(\widetilde{a})|^2~d\widetilde{a}=1$. For plots of the probability densities corresponding to the eigenfunctions, as specified in equations (\ref{zeroeig}), (\ref{sol2}), (\ref{sol3}), (\ref{sol4}) and (\ref{sol5}) across different values of the ordering parameter $p$, we refer to Figure~\ref{figure1geon} and \ref{figure1geon1}.
From the analysis of the previously obtained polynomial solutions, a pattern emerges, suggesting the following quantization condition
\begin{equation}\label{condizione}
\sqrt{\beta}=4(2n+1+p)^2,\quad n=0,1,2,\cdots.    
\end{equation}
Here, $n$ represents the degree of the polynomial solution. By using equation (\ref{coeff}), we can reformulate equation (\ref{condizione}) as
\begin{equation}\label{ilovik}
\rho_0=\frac{9\rho_{Pl}}{64(2n+1+p)},\quad n=0,1,2,\cdots.
\end{equation}
To ensure consistency, we observe that the above result correctly
reproduces equation (\ref{q0}) when $p=0$.  Qualitatively, we also
confirm the behavior of the probability density: the probabilistic
values of the scale factor exhibit more peaks for the preferred values if we go beyond the ground state. Finally, we observe that the quantization condition (\ref{ilovik}) constrains the possible values that the ordering factor may take on. More precisely, we need to require that $p$ is not a negative odd number, that is $p\neq -(2n+1)$ with $n\in\mathbb{N}\cup\{0\}$. 

It is important to note that the conclusions reached in equations \eqref{sol2}, \eqref{sol3}, \eqref{sol4}, and \eqref{psip4} are  subject to the specific choice of operator ambiguities of a polynomial kind, as expressed in equation \eqref{factorord}. If different functions, such as non-polynomial forms, were used for the factor ordering, the resulting WDW equation and its solutions would be significantly different. The redefinition would no longer be guaranteed to be a polynomial in \(a\), leading to potentially different physical interpretations.

\section{Effects of the Cosmological Constant}
Incorporating the cosmological constant into our analysis highlights three key areas worth investigating. Firstly, from a mathematical perspective, the importance of examining the self-adjointness of the Hamiltonian arises. The introduction of the cosmological constant renders the potential unbounded from below, making this examination crucial. Secondly, the issue of quantization persists, especially under conditions where $\Lambda$ is maintained as non-zero. Thirdly, the configuration of the potential now indicates the possibility of a tunneling solution, facilitated by the presence of the cosmological horizon. This scenario mirrors the classical understanding described in \cite{deSitter}, where the effective potential in the geodesic equation for a Schwarzschild-deSitter metric exhibits a local maximum, suggesting a similar tunneling phenomenon.

\subsection{Analysis of the potential}
\begin{figure}[ht!]
\centering
\includegraphics[width=0.4\textwidth]{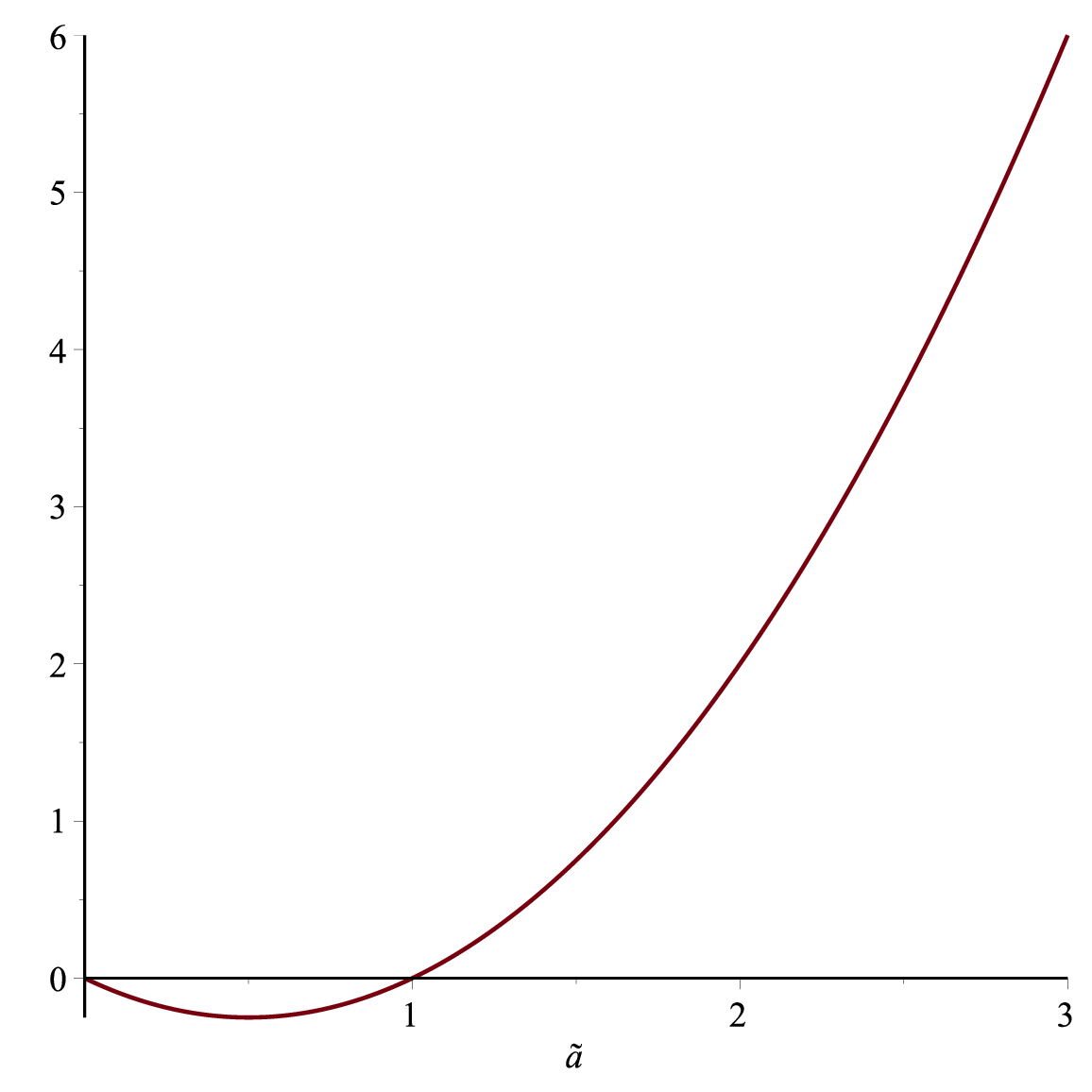}
\includegraphics[width=0.4\textwidth]{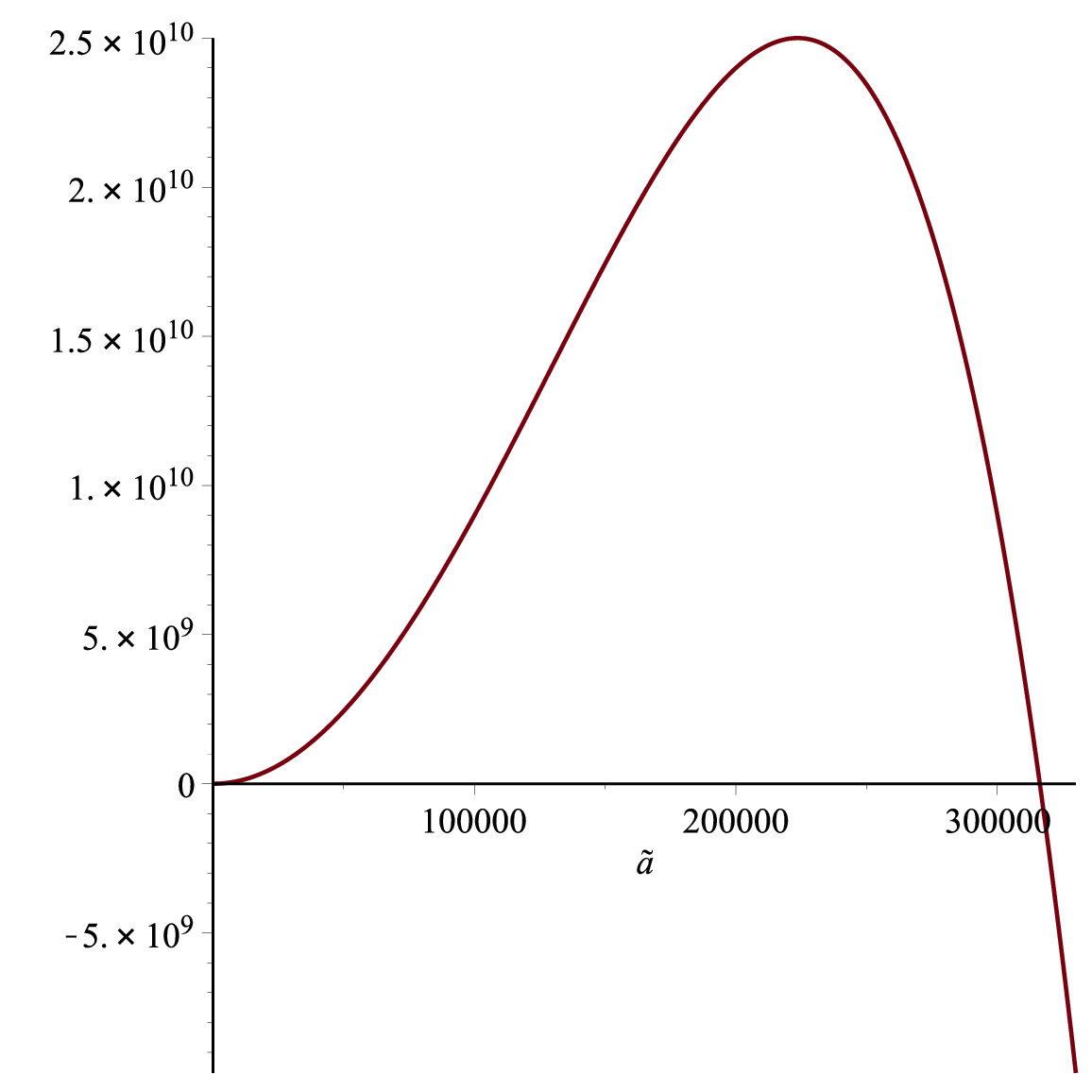}
\caption{
Graphical representation of the effective potential (\ref{pp}) for $\epsilon=10^{-11}$. The potential exhibits a classical forbidden region between $1$ and $3.16\cdot 10^{5}$. The maximum occurs at $2.24\cdot 10^5$.
}
\label{figure1}
\end{figure}

Transitioning to the de Sitter scenario, which is distinguished by a
positive cosmological constant ($\Lambda > 0$), implies a non-zero
vacuum energy density ($\rho_{vac} > 0$). To provide a grasp of the
magnitudes of the coefficients introduced in equation (\ref{coeff}),
let us consider a few illustrative densities. The Planck density,
$\rho_{Pl}$, is notably large at $5.1 \cdot 10^{96}$ Kg/m$^3$. In
stark contrast, the vacuum energy density, $\rho_{vac}$, is
significantly smaller, at $5.9 \cdot 10^{-27}$ Kg/m$^3$. For a typical
cloud core, we anticipate an initial density, $\rho_0$, approximately
$10^{-16}$ Kg/m$^3$ \cite{Bode}. With the densities defined, we
estimate an order of magnitudes using 
\begin{equation}\label{ordering}
\frac{\rho_{vac}}{\rho_0}\ll 1\ll\frac{\rho_{Pl}}{\rho_0},
\end{equation}
where $\rho_{vac}/\rho_0\sim 10^{-11}$ and $\rho_{Pl}/\rho_0\sim 10^{112}$. Consequently, $|\alpha|\sim 10^{214}$ and $\beta\sim 10^{224}$ emerge. This implies that the terms with $\alpha$ in the effective potential become influential only at large values of $\widetilde{a}$. This can be easily seen if we rewrite (\ref{Veff}) as follows
\begin{equation}\label{pp}
\mathcal{V}_{eff}(\widetilde{a})=\frac{U_{eff}(\widetilde{a})}{\beta}=-\epsilon\widetilde{a}^4+\widetilde{a}(\widetilde{a}-1),\quad \epsilon=\frac{\rho_{vac}}{\rho_0}.    
\end{equation}
Then, Figure~\ref{figure1} clearly shows that the effective potential exhibits a minimum between $0$ and $1$ followed by a high maximum for large values of $\widetilde{a}$.
A straightforward application of perturbation theory indicates that the smallest positive intersection $\widetilde{a}_-$ occurs at
\begin{equation}\label{ameno}
\widetilde{a}_-=1+\epsilon+\mathcal{O}(\epsilon^2),
\end{equation}
while the minimum is located at
\begin{equation}\label{minimo}
\widetilde{a}_m=\frac{1}{2}+\frac{\epsilon}{4}+\mathcal{O}(\epsilon^2).    
\end{equation}
For both $\widetilde{a}_-$ and $\widetilde{a}_m$ the first-order corrections prove to be extraordinarily small, approximately of the order of $10^{-11}$. However, for later purposes we also need precise formulae for the positive intersections of the effective potential with the $\widetilde{a}$-axis and the exact locations of the minimum and maximum in the potential. As a first step in this direction we rewrite the effective potential as follows
\begin{equation}\label{rew}
\mathcal{V}_{eff}(\widetilde{a})=-\epsilon\widetilde{a}\left(\widetilde{a}^3+3p\widetilde{a}+2q\right),\quad p=-\frac{1}{3\epsilon},\quad
q=\frac{1}{2\epsilon}.
\end{equation}
Note that the cubic polynomial in equation (\ref{rew}) is already in reduced form. By employing Vieta's theorem, if $\widetilde{a}_i$ (with $i=1,2,3$) denotes the roots of the reduced cubic, we deduce $\widetilde{a}_1+\widetilde{a}_2+\widetilde{a}_3=0$ and $\widetilde{a}_1 \widetilde{a}_2 \widetilde{a}_3=-2q$. Given that $q>0$, if the cubic has three real roots, there can only be either one or two negative roots. The number of real solutions for our cubic equation is determined by the sign of the discriminant
\begin{equation}
D=q^2+p^3=\frac{1}{27\epsilon^3}\left(\frac{27}{4}\epsilon-1\right).    
\end{equation}
Since $\epsilon\ll 1$, we are clearly in the regime $\epsilon<4/27$
and we immediately conclude that $D<0$. This implies that the cubic in
(\ref{rew}) has three real roots. We must make two important
observations at this stage. Firstly, in the current regime, aside from
$\widetilde{a}=0$, the effective potential reveals two positive roots,
henceforth referred to as $a_\pm$. Subsequently, we will rename
$\widetilde{a}_1$ as $\widetilde{a}_{-}$. It is noteworthy that the
potential displays both a minimum and maximum value, which we will
calculate shortly. In analogy with the standard Schr\"odinger equation, we observe a classically forbidden region $\widetilde{a}_{-}< \widetilde{a} < \widetilde{a}_+$ and two classically allowed regions, specifically when $0<\widetilde{a}<\widetilde{a}_-$ and $\widetilde{a}>\widetilde{a}_+$.  As detailed in \cite{Bron}, we define 
\begin{equation}
\mathfrak{r}=\frac{1}{\sqrt{3\epsilon}},\quad
\cos{\phi}=\frac{3}{2}\sqrt{3\epsilon}. 
\end{equation}
Recalling that $\epsilon\ll 1$ allows to conclude that $\phi$ lies within the range $(0,\pi/2)$. With this observation, the two positive roots can be expressed as
\begin{equation}\label{apm}
\widetilde{a}_\pm=\frac{2}{\sqrt{3\epsilon}}\cos{\left(\frac{\pi\mp\phi}{3}\right)},\quad
\phi=\cos^{-1}{\left(\frac{3}{2}\sqrt{3\epsilon}\right)}.
\end{equation}
Since $(\pi-\phi)/3$ and $(\pi+\phi)/3$ fall within the ranges $\pi/6<(\pi-\phi)/3<\pi/3$ and $\pi/3<(\pi+\phi)/3<\pi/2$, respectively, we infer that $\widetilde{a}_+>\widetilde{a}_-$. Taking into account that $\widetilde{a}_+$ admits the perturbative expansion
\begin{equation}\label{aplus}
\widetilde{a}_+=\frac{1}{\sqrt{\epsilon}}+\mathcal{O}(1), 
\end{equation}
the separation between these turning points, denoted as $\Delta \widetilde{a}_\pm=\widetilde{a}_+-\widetilde{a}_-$, can be calculated for the case of $\epsilon\ll 1$ as
\begin{equation}\label{width}
\Delta \widetilde{a}_\pm=\frac{1}{\sqrt{\epsilon}}+\mathcal{O}(1).
\end{equation}
In order to compute the maximum and minimum displayed in Figure~\ref{figure1} for the case $\epsilon\ll 1$, we impose that $d\mathcal{V}_{eff}/d\widetilde{a}=0$. This condition gives rise to the following reduced cubic
\begin{equation}
\widetilde{a}^3+3\mathfrak{p}\widetilde{a}+2\mathfrak{q}=0,\quad
\mathfrak{p}=-\frac{1}{6\epsilon},\quad
\mathfrak{q}=\frac{1}{8\epsilon}.
\end{equation}
The corresponding discriminant
\begin{equation}
\mathfrak{D}=\frac{1}{216\epsilon^3}\left(\frac{27}{8}\epsilon-1\right)  
\end{equation}
is clearly negative since $\epsilon\ll 1$. As a consequence, the cubic above admits three real roots. Following \cite{Bron}, we define
\begin{equation}
\widehat{\mathfrak{r}}=\frac{1}{\sqrt{6\epsilon}},\quad 
\cos{\widehat{\phi}}=\frac{3}{4}\sqrt{6\epsilon},
\end{equation}
by means of which the two positive roots can be expressed as
\begin{equation}\label{bubu}
\widetilde{a}_M=\frac{2}{\sqrt{6\epsilon}}\cos{\left(\frac{\pi-\widehat{\phi}}{3}\right)},\quad
\widetilde{a}_m=\frac{2}{\sqrt{6\epsilon}}\cos{\left(\frac{\pi+\widehat{\phi}}{3}\right)},\quad
\widehat{\phi}=\cos^{-1}{\left(\frac{3}{4}\sqrt{6\epsilon}\right)}.
\end{equation}
A consistency check shows that the perturbation expansion for $\widetilde{a}_m$ given by (\ref{bubu}) coincides with (\ref{minimo}) while the corresponding expansion for the maximum is given by
\begin{equation}\label{locaM}
\widetilde{a}_M=\frac{1}{\sqrt{2\epsilon}}+\mathcal{O}(1).    
\end{equation}
It is gratifying to observe that $\widetilde{a}_M\to+\infty$ as $\epsilon\to 0$ which correctly signals that the effective potential has a global minimum when $\epsilon=0$. Taking into account that $\epsilon \ll  1$, the effective potential at these points is given by
\begin{equation}
 \mathcal{V}_{eff}(\widetilde{a}_m)=-\frac{1}{4}-\frac{\epsilon}{16}+\mathcal{O}(\epsilon^2),\quad
\mathcal{V}_{eff}(\widetilde{a}_M)=\frac{1}{4\epsilon}+\mathcal{O}\left(\frac{1}{\sqrt{\epsilon}}\right).\label{VaM}
\end{equation}
It follows that the height of the barrier is
\begin{equation}
    \Delta \mathcal{V}_{eff}=\mathcal{V}_{eff}(\widetilde{a}_M)-\mathcal{V}_{eff}(\widetilde{a}_m)=\frac{1}{4\epsilon}+\mathcal{O}\left(\frac{1}{\sqrt{\epsilon}}\right),
\end{equation}
By means of (\ref{pp}) we can switch back to the effective potential $U_{eff}$ and find that
\begin{equation}
\Delta U_{eff}\sim\frac{\beta}{4\epsilon}=\frac{81\rho_{Pl}^2}{1024\rho_0\rho_{vac}}\sim 7.9\cdot 10^{223}.
\end{equation}

\subsection{Spectral considerations on the Hamiltonian}
When dealing with differential operators, especially in quantum
mechanics, we often want to ensure that the operator is
self-adjoint. This is crucial because self-adjoint operators have real
eigenvalues, which correspond to physically observable
quantities. However, not all differential operators possess this
property. Their self-adjointness can depend on the behavior of
solutions to the associated differential equation at certain endpoints
of an interval. Our variable $a$ varies in the domain $(0, \infty)$ and
  not on the whole real line.  Therefore, the spectral properties of our
  problem are connected to the non self-adjointness of the radial momentum
  operator in standard quantum mechanics \cite{Gil}.

Given a Sturm-Liouville type operator
\begin{equation}\label{SLE}
Ly=-\frac{d^2 y}{dx^2}+U(x)y,\quad 0\leq x<+\infty,
\end{equation}
where $U$ is the potential, the behavior of solutions at the endpoints can be classified into two main categories: the limit point case (LP) and the limit circle case (LC). 
We say that (\ref{SLE}) is in the LP case, if there is only one linearly independent solution to the differential equation that is square-integrable as $x$ approaches the boundary. In this scenario, the operator naturally exhibits self-adjointness across a broad spectrum of boundary conditions. On the other hand, (\ref{SLE}) is in the LC case, whenever there are two linearly independent solutions to the differential equation and both are square-integrable as $x\to+\infty$. In this scenario, since self-adjointness is not guaranteed, one must choose appropriate boundary conditions. More precisely, there can be a whole family of possible self-adjoint extensions, each corresponding to a different boundary condition. The distinction between the LP and LC cases is crucial. In quantum mechanics, a self-adjoint Hamiltonian ensures that the energy eigenvalues are real and that the time evolution of a system is unitary. If an operator is in the LC case, and we do not handle it correctly, we might end up with a non-self-adjoint Hamiltonian, leading to non-physical results. In our present problem, we observe that due to the fact that the effective potential (\ref{Veff}) is unbounded from below, the Hamiltonian operator
\begin{equation}\label{Ham}
H=-\frac{d^2}{d\widetilde{a}^2}+U_{eff}(\widetilde{a}),\quad
U_{eff}(\widetilde{a})=\alpha \widetilde{a}^4+\beta\widetilde{a}(\widetilde{a}-1)
\end{equation}
with $\alpha$ and $\beta$ defined as in (\ref{coeff}) is not essentially self-adjoint. Nevertheless, we will soon show that the Schr\"{o}dinger equation (\ref{WDWfin}) always admits a square integrable solution on the interval $[0,\infty)$. Moreover, since $U_{eff}(\widetilde{a})\geq 0$ on the interval $[\widetilde{a}_-,\widetilde{a}_+]$, the first corollary after Theorem~$3.3$ in \cite{Berezin} predicts that such a solution has at most one zero on the aforementioned interval. In order to convince ourselves that the Hamiltonian $H$ defined in (\ref{Ham}) is in the limit circle case, we need to study the behavior of the solution in the regimes $\widetilde{a}\to\infty$ and $\widetilde{a}\to 0$. Firstly, we observe that our (\ref{Ham}) is a special case of (4.1) and (4.2) in \cite{Berezin} with $\kappa=0$. Moreover, since the effective potential is asymptotically divergent, we are in case c) with $f(\widetilde{a})=\sqrt{-U_{eff}(\widetilde{a})}$ (see page 84 in \cite{Berezin}). In order to apply Theorem~$4.6$ therein, we need to check if the effective potential satisfies the conditions 
\begin{equation}\label{condizioni}
\int_{\widetilde{a}_0}^\infty\frac{|U_{eff}^{'}(\widetilde{a})|^2}{|U_{eff}(\widetilde{a})|^{5/2}}d\widetilde{a}<\infty,\quad
\int_{\widetilde{a}_0}^\infty\frac{|U_{eff}^{''}(\widetilde{a})|^2}{|U_{eff}(\widetilde{a})|^{3/2}}d\widetilde{a}<\infty,
\end{equation}
where $\widetilde{a}_0$ in the above integrals can be chosen arbitrarily large. Here, a prime denotes differentiation with respect to $\widetilde{a}$. Let us pick $\widetilde{a}_0\gg 1$ so that we only need to consider the asymptotic behavior of the integrands above. A routine calculation shows that
\begin{equation}
    \frac{|U_{eff}^{'}(\widetilde{a})|^2}{|U_{eff}(\widetilde{a})|^{5/2}}=\frac{16}{\sqrt{\beta\epsilon}\widetilde{a}^4}+\mathcal{O}\left(\frac{1}{\widetilde{a}^5}\right),\quad
    \frac{|U_{eff}^{''}(\widetilde{a})|^2}{|U_{eff}(\widetilde{a})|^{3/2}}=\frac{12}{\sqrt{\beta\epsilon}\widetilde{a}^4}+\mathcal{O}\left(\frac{1}{\widetilde{a}^6}\right)
\end{equation}
and we immediately realize that both conditions appearing in (\ref{condizioni}) are fulfilled. Then, Theorem~$4.6$ implies that (\ref{WDWfin}) admits a pair of solutions with the following asymptotic behavior as $\widetilde{a}\to\infty$
\begin{equation}
\Psi_{\pm,\infty}(\widetilde{a})=\frac{1}{\sqrt[4]{-U_{eff}(\widetilde{a})}}\mbox{exp}\left(\pm i \int_{\widetilde{a}_0}^{\widetilde{a}}\sqrt{-U_{eff}(s)}ds\right)\left[1+\mathcal{O}(1)\right].    
\end{equation}
If we take into account that $\widetilde{a}_0\gg 1$, we can further approximate the above formula as follows
\begin{equation}\label{apr1}
\Psi_{\pm,\infty}(\widetilde{a})=\frac{1}{\widetilde{a}}e^{\pm \frac{i}{3}\sqrt{\beta\epsilon}\widetilde{a}^3}\left[1+\mathcal{O}(1)\right],
\end{equation}
which is reminiscent of (5.13) in \cite{Hartle}. Note that the second factor in (\ref{apr1}) oscillates and is bounded. Hence, both solutions have integrable squares on the interval $[\widetilde{a}_0,\infty)$ even though the original Hamiltonian is not essentially self-adjoint. In order to investigate the solutions in a neighborhood of $\widetilde{a}=0$, we first observe that in the limit of $\widetilde{a}\to 0$ (\ref{WDWfin}) becomes
\begin{equation}
\frac{d^2\Psi_0}{d\widetilde{a}^2}+\beta\widetilde{a}\Psi(\widetilde{a})=0.    
\end{equation}
Its solution can be expressed in terms of Airy functions as follows
\begin{equation}
   \Psi_0(\widetilde{a})=c_1\mbox{Ai}(-\sqrt[3]{\beta}\widetilde{a})+c_2\mbox{Bi}(-\sqrt[3]{\beta}\widetilde{a}).
\end{equation}
Note that both Airy functions are finite at $\widetilde{a}=0$ and square integrable on the interval $[0,c]$ with $c\ll 1$. Hence, according to theorem D in \cite{McLeod}, we conclude that the differential operator (\ref{Ham}) is in the limit circle case. It is also worth emphasizing that potentials unbounded from below are not merely theoretical artifacts. They have practical relevance in high-energy physics, particularly in the context of field localization on the 3-brane within braneworld models featuring warped extra dimensions \cite{seven}. As highlighted in \cite{four}, these types of potentials can be approximated with a significant degree of accuracy in real-world mesoscopic systems. 

The fact that equation (\ref{Ham}) falls within the limit circle case
suggests that the operator is not self-adjoint unless additional
boundary conditions are introduced or a suitable domain of definition
for the operator is specified. 

  However, for those magnitudes of the parameter $\epsilon$ associated with various black hole scenarios (refer to Table~\ref{tableEins}), it can be demonstrated that our Hamiltonian acquires Parity-Time (PT) symmetry in the manner described by \cite{Bender}. It is inherently time-invariant, as time does not feature in the WDW equation. To verify the parity symmetry across the $\epsilon$ ranges outlined in Table~\ref{tableEins}, introducing a change of variable $\widetilde{a}=x+\widetilde{a}_m$, accompanied by a vertical shift by $\mathcal{V}_{eff}(\widetilde{a}_M)$, proves useful. Finally, with the help of  (\ref{minimo}), (\ref{locaM}), and (\ref{VaM}), we can rewrite equation (\ref{WDWfin}) as
\begin{equation}
    -\frac{d\psi}{dx^2}+\beta\left[-\epsilon x^4+x^2-\frac{1}{4\epsilon}\right]\psi(x)=0.
\end{equation}
It is straightforward to verify that the operator above maintains its invariance under the transformation $x \rightarrow -x$. With an appropriately defined inner product, the eigenfunctions of a PT-symmetric Hamiltonian possess positive norms and demonstrate unitary time evolution \cite{Bender1}. These characteristics are essential for the consistency of quantum theories. Furthermore, \cite{Bender1} demonstrated that all eigenvalues of a PT-symmetric Hamiltonian with an inverted quartic potential are real and positive. This property is corroborated by our findings in the following subsection, where we establish a quantization condition for the density $\rho_0$, expressed in relation to the Planck density.

Last but not least, an intriguing observation from equation (\ref{apr1}) hints at a form of tunneling. Despite the probability density $|\Psi_{\pm,\infty}|^2\sim \widetilde{a}^{-2}$ decaying as $\widetilde{a}\to\infty$, it indicates a non-negligible probability in the region $(\widetilde{a}_+,\infty)$. To delve deeper into this phenomenon, let us first observe that in the $\epsilon\to 0$ limit, equation (\ref{aplus}) which depicts the most significant intersection of the effective potential with the $\widetilde{a}$-axis, simplifies to
\begin{equation}\label{radius}
\widetilde{a}_+=\frac{1}{\sqrt{\epsilon}}=\sqrt{\frac{3k}{\Lambda}},
\end{equation}
where we employed (\ref{pp}) and $k$ is defined via (\ref{kappa}). This result implies that the cosmological horizon is adjusted by a factor of $\sqrt{k}$. Consequently, $\widetilde{a}_+$ can be understood as the cosmological horizon in our context. To understand the behavior of (\ref{radius}) in the \(\Lambda \to 0\) limit, we focus on its dependence on \(\Lambda\). First of all, this equation represents the cosmological horizon, where \(k\) is related to the initial density \(\rho_0\) of the collapsing dust cloud. As \(\Lambda \to 0\), the term \(\frac{\sqrt{3k}}{\Lambda}\) diverges. This implies that the cosmological horizon \(a_+\) moves to infinity. Physically, this is consistent with the classical picture where, in the absence of a cosmological constant, there is no cosmological horizon, and the potential barrier introduced by \(\Lambda\) vanishes.

\subsection{An approximated quantization condition}
An approximated quantization condition can be derived by studying the differential equation (\ref{WDWfin}) in a neighborhood of the minimum in the effective potential. To this purpose, we can add and subtract the positive term $-U_{eff}(\widetilde{a}_m)$ to the effective potential and recast the WDW equation in the equivalent form
\begin{equation}\label{ODE01}
-\frac{d^2\Psi}{d\widetilde{a}^2}+\left[U_{eff}(\widetilde{a})-U_{eff}(\widetilde{a}_m)\right]\Psi(\widetilde{a})=-U_{eff}(\widetilde{a}_m)\Psi(\widetilde{a}).
\end{equation}
Through this process, we have effectively shifted the potential upward by $-U_{eff}(\widetilde{a}_m)$. This newly introduced term now serves as a positive spectral parameter. Taking into account that $\widetilde{a}_m$ satisfies the equation $dU_{eff}/d\widetilde{a}=0$ and performing a Taylor expansion around $\widetilde{a}=\widetilde{a}_m$, it is not difficult to verify that
\begin{equation}
U_{eff}(\widetilde{a})-U_{eff}(\widetilde{a}_m)=\beta\left[-\epsilon(\widetilde{a}-\widetilde{a}_m)^4-4\epsilon\widetilde{a}_m(\widetilde{a}-\widetilde{a}_m)^3+(1-6\epsilon\widetilde{a}_m^2)(\widetilde{a}-\widetilde{a}_m)^2\right].
\end{equation}
Finally, by means of the coordinate change $\widehat{a}=\widetilde{a}-\widetilde{a}_m$ we can rewrite (\ref{ODE01}) as follows
\begin{equation}\label{TODE}
-\frac{d^2\Psi}{d\widehat{a}^2}+\mathcal{U}_{eff}(\widehat{a})\Psi(\widehat{a})=-U_{eff}(\widetilde{a}_m)\Psi(\widehat{a}),\quad\widehat{a}\geq-\widetilde{a}_m
\end{equation}
with
\begin{equation}
    \mathcal{U}_{eff}(\widehat{a})=
    \beta\left[(1-6\epsilon\widetilde{a}_m^2)\widehat{a}^2-4\epsilon\widetilde{a}_m\widehat{a}^3-\epsilon\widehat{a}^4\right].
\end{equation}
In a neighborhood of $\widehat{a}=0$, (\ref{TODE}) becomes the equation of a harmonic oscillator, namely
\begin{equation}\label{ODE02}
    -\frac{d^2\Psi_o}{d\widehat{a}^2}+K\widehat{a}^2\Psi_o(\widehat{a})=E\Psi_o(\widehat{a}),\quad
    K=\beta(1-6\epsilon\widetilde{a}_m^2),\quad
    E=-U_{eff}(\widetilde{a}_m).
\end{equation}
Since $K$ is positive as it can be seen from the following expansion in the small parameter $\epsilon$
\begin{equation}
    K=\beta\left[1-\frac{3}{2}\epsilon+\mathcal{O}(\epsilon^2)\right],
\end{equation}
we can proceed as in \cite{Schiff}. Namely, we introduce the new variable $\xi=\sqrt[4]{K}\widehat{a}$ and cast (\ref{ODE02}) into the form
\begin{equation}
    \frac{d^2\Psi_o}{d\xi^2}+(\lambda-\xi^2)\Psi_o(\xi)=0,\quad
    \lambda=\frac{E}{\sqrt{K}}.
\end{equation}
Then, the quantization condition reads \cite{Schiff}
\begin{equation}
    2n+1=-\frac{\mathcal{U}_{eff}(\widehat{a}_m)}{\sqrt{\beta(1-6\epsilon\widetilde{a}_m^2)}},\quad n=0,1,2,\cdots.
\end{equation}
Expanding the r.h.s. of the above expression in the small parameter $\epsilon$ yields
\begin{equation}\label{quan1}
   2n+1=\frac{\sqrt{\beta}}{4}\left[1+\epsilon+\mathcal{O}(\epsilon^2)\right],\quad n=0,1,2,\cdots.
\end{equation}
As a consistency check for the above procedure, we observe that in the case $\epsilon=0$ the above formula correctly reproduces the quantization condition obtained in \cite{ourpaper}. Finally, with the help of (\ref{coeff}) we can rewrite (\ref{quan1}) as
\begin{equation}\label{ac}
   16(2n+1)^2= \frac{81}{256}\left(\frac{\rho_{Pl}}{\rho_0}\right)^2\left[1+\frac{2\rho_{vac}}{\rho_0}+\mathcal{O}\left(\frac{\rho_{vac}^2}{\rho_0^2}\right)\right],\quad n=0,1,2,\cdots.
\end{equation}
The approximated quantization condition derived in this section highlights the discrete nature of the density \(\rho_0\) in the presence of a positive cosmological constant \(\Lambda\). By expanding the effective potential around its minimum and drawing an analogy to a quantum harmonic oscillator, we obtain (\ref{ac}). This condition shows that \(\rho_0\) is quantized, with levels influenced by the Planck density and the vacuum energy density, reflecting the impact of quantum gravitational effects.

\subsection{WKB analysis and tunneling probability}
\begin{figure}[ht!]
\centering
\includegraphics[width=0.4\textwidth]{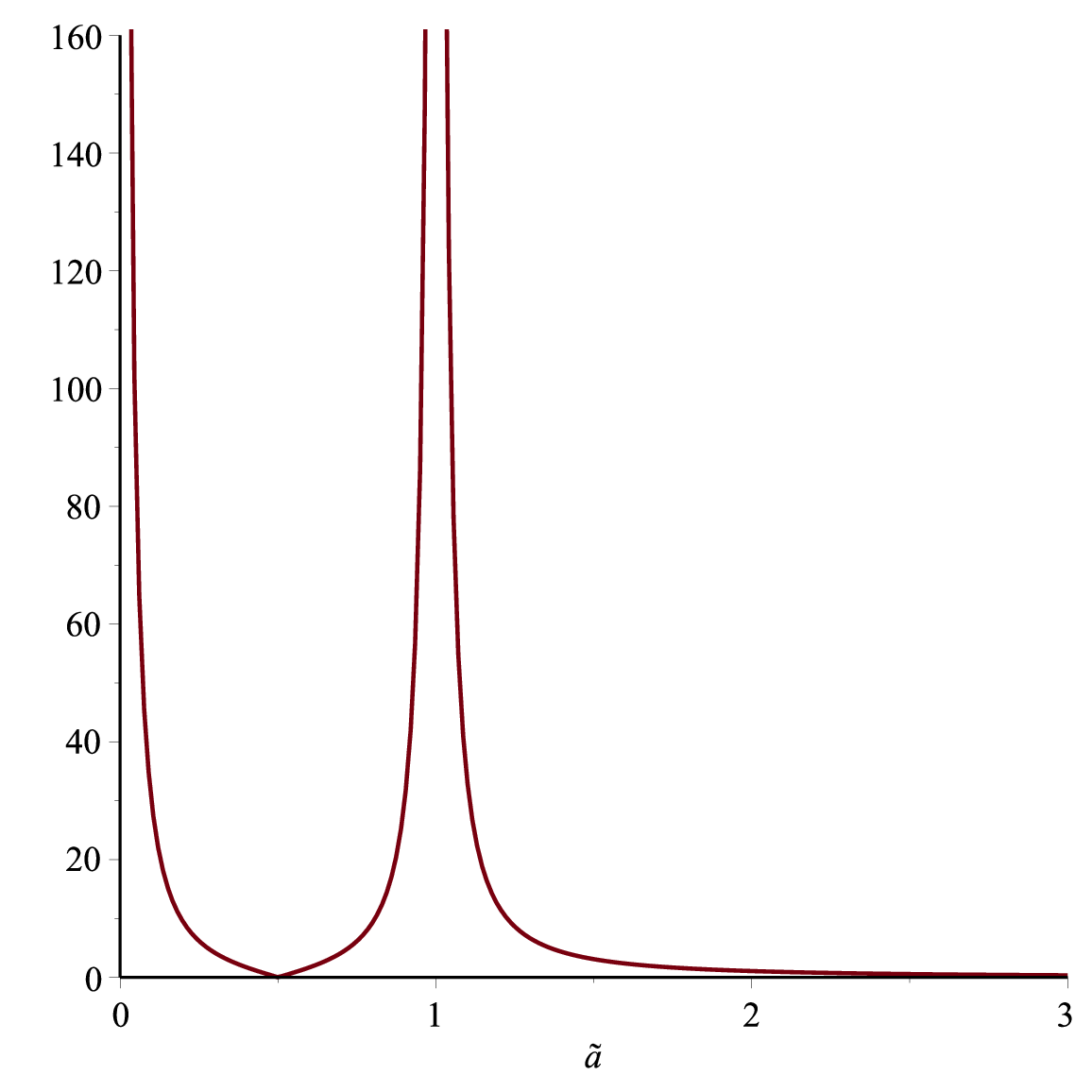}
\includegraphics[width=0.4\textwidth]{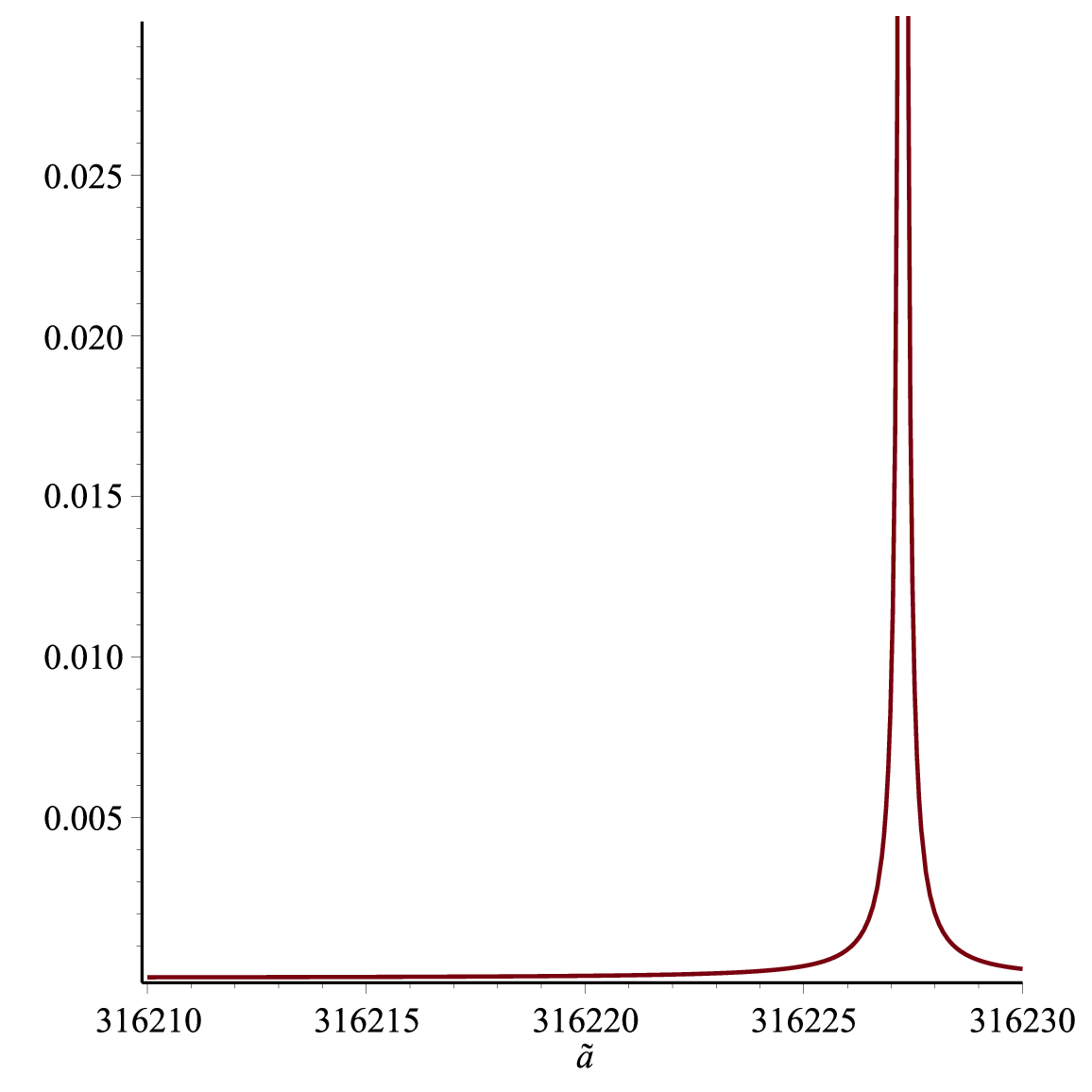}
\caption{
Plot of $|V^{'}/V^{3/2}|$ for $\epsilon=10^{-11}$ and $V$ defined as in (\ref{pora}). In view of the condition (\ref{diciasette}), the WKB approximation holds exceptionally well everywhere modulo some tiny intervals centered at the turning points of the potential.
}
\label{figureWKB}
\end{figure}

Let us first rewrite the Schr\"{o}dinger equation (\ref{WDWfin}) as follows
\begin{equation}\label{pora}
-\mu^2\frac{d^2\Psi}{d\widetilde{a}^2}+V(\widetilde{a})\Psi(\widetilde{a})=0,\quad
V(\widetilde{a})=-\widetilde{a}+\widetilde{a}^2-\epsilon\widetilde{a}^4,\quad
\mu^2=\frac{1}{\beta},\quad \widetilde{a}\geq 0.
\end{equation}
We recall that this is a one-dimensional Schr\"{o}dinger equation for a particle with zero total energy and half the unit mass. Note that according to our previous estimates $\mu\sim 10^{-112}$ and $\epsilon\sim 10^{-11}$. This observation suggests that we can construct a WKB solution in terms of the small parameter $\mu$. Let us introduce the local momentum $p^2(\widetilde{a})=-V(\widetilde{a})$ which in turn allows defining a local de Broglie wavelength
\begin{equation}\label{zero}
\lambda(\widetilde{a})=\frac{\mu}{p(\widetilde{a})}.   
\end{equation}
In view of the notation introduced above, it is convenient to rewrite (\ref{pora}) as follows
\begin{equation}\label{SCHR}
   -\mu^2\frac{d^2\Psi}{d\widetilde{a}^2}=p^2(\widetilde{a})\Psi(\widetilde{a}),\quad
   \widetilde{a}\geq 0.
\end{equation}
The potential admits three physical turning points at
$\widetilde{a}=0$ and $\widetilde{a}_\pm$ which are given by
(\ref{ameno}) and (\ref{aplus}), respectively.  Again, drawing an
  analogy to the standard Schr\"odinger equation, we can say that there are two classically allowed regions where $V(\widetilde{a})<0$ and $p^2(\widetilde{a})>0$. The first region denoted by the symbol I occurs for $\widetilde{a}\in (0,\widetilde{a}_{-})$ while the second region called III extends over the interval $(\widetilde{a}_+,\infty)$. In the classically forbidden region, we have $V(\widetilde{a})>0$ and thus $p^2(\widetilde{a})<0$. We use the symbol II to denote such region, which is limited to the interval $(\widetilde{a}_{-},\widetilde{a}_+)$. If we write the wave function as 
\begin{equation}
    \Psi(\widetilde{a})=\sqrt{\rho(\widetilde{a})}e^{\frac{i}{\mu}\mathcal{S}(\widetilde{a})},
\end{equation}
where $\rho$ is the probability density, it is straightforward to verify that
\begin{equation}
    \Psi^{*}(\widetilde{a})\frac{d\Psi}{d\widetilde{a}}=\frac{1}{2}\frac{d\rho}{d\widetilde{a}}+\frac{i}{\mu}\rho(\widetilde{a})\frac{d\mathcal{S}}{d\widetilde{a}}
\end{equation}
and therefore, the only non-zero current component is the one along the $\widetilde{a}$-axis and is given by the formula
\begin{equation}\label{unob}
    J_{\widetilde{a}}=2\mu\Im{\left(\Psi^{*}(\widetilde{a})\frac{d\Psi}{d\widetilde{a}}\right)}=2\rho(\widetilde{a})\frac{d\mathcal{S}}{d\widetilde{a}}.
\end{equation}
We recall that in classical physics a fluid with density $\rho(x)$ moving with velocity $v(x)$ has a current density $\rho v=2\rho p$ which contrasted with (\ref{unob})  leads to the conclusion
\begin{equation}\label{duea}
    p(x)\sim\frac{d\mathcal{S}}{d\widetilde{a}}.
\end{equation}
Observe that the equality does not exactly hold in the above formula in general unless we are in the WKB regime. In our WKB approximation scheme, it is more convenient to introduce the ansatz
\begin{equation}\label{ansatz}
    \Psi(\widetilde{a})=e^{\frac{i}{\mu}S(\widetilde{a})},
\end{equation}
where $\Re{(S)}$ and $\Im{(S)}$ represent the phase and magnitude of the wave function. By replacing (\ref{ansatz}) into (\ref{SCHR}), we obtain the following nonlinear differential equation for $S$
\begin{equation}\label{tre}
\left(\frac{dS}{d\widetilde{a}}\right)^2-i\mu\frac{d^2 S}{d\widetilde{a}^2}=p^2(\widetilde{a}).    
\end{equation}
If we take $\mu$ to be the small parameter in a systematic expansion of $S$ of the form
\begin{equation}\label{espansione}
    S(\widetilde{a})=S_0(\widetilde{a})+\mu S_1(\widetilde{a})+\mathcal{O}(\mu^2),
\end{equation}
replacing (\ref{espansione}) into (\ref{tre}) and neglecting terms of order $\mu^2$ and higher lead to the equation
\begin{equation}\label{cinque}
    \left[\left(\frac{dS_0}{d\widetilde{a}}\right)^2-p^2(\widetilde{a})\right]+\mu\left(2\frac{dS_0}{d\widetilde{a}}\frac{dS_1}{d\widetilde{a}}-i\frac{d^2 S_0}{d\widetilde{a}^2}\right)+\mathcal{O}(\mu^2)=0.
\end{equation}
The above equation is satisfied provided that
\begin{eqnarray}
    \left(\frac{dS_0}{d\widetilde{a}}\right)^2&=&p^2(\widetilde{a}),\label{sei}\\
    \frac{dS_1}{d\widetilde{a}}&=&\frac{i}{2}\frac{d^2 S_0/d\widetilde{a}^2}{dS_0/d\widetilde{a}}.\label{sette}
\end{eqnarray}
Equation (\ref{sei}) can be immediately integrated with solutions
\begin{equation}\label{otto}
    S_0(\widetilde{a})=\pm\int_{\widetilde{a}_0}^{\widetilde{a}}p(s)ds,
\end{equation}
where $\widetilde{a}_0$ is an initial point to be accordingly adjusted before performing the integral. Moreover, with the help of (\ref{sei}) equation (\ref{sette}) can be rewritten as
\begin{equation}\label{ottoa}
    \frac{dS_1}{d\widetilde{a}}=\frac{i}{2}\frac{d}{d\widetilde{a}}\ln{p(\widetilde{a})}.
\end{equation}
This is readily solved to give
\begin{equation}\label{nove}
    iS_1(\widetilde{a})=-\frac{1}{2}\ln{p(\widetilde{a})}+C^{'}.
\end{equation}
Let us now reconstruct the wave function to this order of approximation by means of (\ref{otto}) and (\ref{nove}). We find that
\begin{equation}\label{dieci}
    \Psi(\widetilde{a})=\frac{A}{\sqrt{p(\widetilde{a})}}e^{\pm \frac{i}{\mu}\int_{\widetilde{a}_0}^{\widetilde{a}}p(s)ds}.
\end{equation}
This is the basic solution in the WKB approximation. We do not attempt to normalize this wave function because the region of validity of this approximation is still unclear and will be discussed in this section. Let us recall that the observables for the basic solution are
\begin{enumerate}
    \item 
    the probability density given by
\begin{equation}\label{undici}
    \rho(\widetilde{a})=|\Psi(\widetilde{a})|^2=\frac{|A|^2}{p(\widetilde{a})},
\end{equation}
    where we used (\ref{dieci}). Clearly, $\rho$ is higher when $p$ is small in those regions.
    \item 
    The probability current according to (\ref{unob}) coupled with (\ref{duea}) and (\ref{undici}) simplifies to
    \begin{equation}\label{dodici}
        J_{\widetilde{a}}=2|A|^2.
    \end{equation}
    Clearly, the current is constant. This is expected because a position dependent current for the zero energy eigenstate is not possible as it would violate the current conservation equation $\partial_{\widetilde{a}}J_{\widetilde{a}}+\partial_t\rho=0$, given that $\rho$ is time-independent. For a discussion of the probability current in the context of the WDW equation and its implications for quantum cosmology, we refer to the seminal papers \cite{WDW,Vil1986}.

\end{enumerate}
Using (\ref{dieci}), we can finally write the general solutions that apply to classically allowed and classically forbidden regions. In the allowed regions I and III, we have $-V(\widetilde{a})>0$ with $p^2(\widetilde{a})=-V(\widetilde{a})$ and hence, the general solution is a superposition of two basic solutions with waves propagating in opposite direction, namely
\begin{eqnarray}
    \Psi_I(\widetilde{a})&=&\frac{A_I}{\sqrt[4]{-V(\widetilde{a})}}e^{\frac{i}{\mu}\int_{\widetilde{a}_{0,I}}^{\widetilde{a}}\sqrt{-V(s)}ds}+\frac{B_I}{\sqrt[4]{-V(\widetilde{a})}}e^{-\frac{i}{\mu}\int_{\widetilde{a}_{0,I}}^{\widetilde{a}}\sqrt{-V(s)}ds},\label{tredici}\\
    \Psi_{III}(\widetilde{a})&=&\frac{A_{III}}{\sqrt[4]{-V(\widetilde{a})}}e^{\frac{i}{\mu}\int_{\widetilde{a}_{0,III}}^{\widetilde{a}}\sqrt{-V(s)}ds}+\frac{B_{III}}{\sqrt[4]{-V(\widetilde{a})}}e^{-\frac{i}{\mu}\int_{\widetilde{a}_{0,III}}^{\widetilde{a}}\sqrt{-V(s)}ds},\label{quattordici}.
\end{eqnarray}
Note that the waves with coefficients $A_I$ and $A_{III}$ move to the right while the second waves with coefficients $B_I$ and $B_{III}$ move to the left. On the forbidden region II the potential is positive and $p^2(\widetilde{a})=-V(\widetilde{a})$, so the general solution reads
\begin{equation}\label{quindici}
    \Psi_{II}(\widetilde{a})=\frac{A_{II}}{\sqrt[4]{V(\widetilde{a})}}e^{\frac{1}{\mu}\int_{\widetilde{a}_{0,II}}^{\widetilde{a}}\sqrt{V(s)}ds}+\frac{B_{II}}{\sqrt[4]{V(\widetilde{a})}}e^{-\frac{1}{\mu}\int_{\widetilde{a}_{0,II}}^{\widetilde{a}}\sqrt{V(s)}ds}.
\end{equation}
Since in the expressions (\ref{tredici})-(\ref{quindici}) the potential vanishes at the turning point, we need to discuss the validity region of our WKB approximation. To this purpose, it is convenient to go back to the expansion 
(\ref{cinque}). We must have that the terms of order $\mu$ in the differential equation are much smaller in magnitude than the $\mathcal{O}(1)$ terms. At each of these orders, we have two terms that are set equal to each other by the differential equations (\ref{sei}) and (\ref{sette}). Hence, it suffices to check that one of the $\mathcal{O}(\mu)$ terms is much smaller than one of the $\mathcal{O}(1)$ terms. For example, we must have $|\mu S_0^{'} S_1^{'}|\ll|S_0^{'}|^2$ where the prime denotes differentiation with respect to $\widetilde{a}$. Since $|S_0^{'}|=|p|$, we obtain $|\mu S_1^{'}|\ll|p|$. By means of (\ref{ottoa}), we note that $|S_1^{'}|=|p^{'}/2p|$ and therefore, $|\mu p^{'}/p|\ll|p|$. With the help of (\ref{zero}) we get, $|\lambda p^{'}|\ll|p|$ which says that the changes in the local momentum over a distance equal to the de Broglie wavelength are small compared to the momentum. Alternatively, we can rewrite the latter condition as
\begin{equation}
    \left|\frac{d\lambda}{d\widetilde{a}}\right|\ll 1,
\end{equation}
that is the local de Broglie wavelength must vary slowly. In our case, we find with the help of (\ref{zero}) that
\begin{equation}
    \left|\frac{d}{d\widetilde{a}}\left(\frac{1}{p}\right)\right|\ll\frac{1}{\mu}=\sqrt{\beta}\sim 10^{112}.
\end{equation}
Since the left-hand side of the above expression can be expressed in terms of the potential as follows
\begin{equation}
    \left|\frac{d}{d\widetilde{a}}\left(\frac{1}{p}\right)\right|=\left|\frac{p^{'}}{p^2}\right|=\left|\frac{V^{'}}{V^{3/2}}\right|
\end{equation}
we conclude that our WKB approximation will hold at all those point on the positive $\widetilde{a}$-axis such that
\begin{equation}\label{diciasette}
    \left|\frac{V^{'}}{V^{3/2}}\right|\ll\frac{1}{\mu}.
\end{equation}
As can be seen in Figure~\ref{figureWKB}, the condition (\ref{diciasette}) is violated only when the object lingers in a region extremely close to the turning points of the potential. To understand how close we can go to the point $\widetilde{a}=0$, we can go back to (\ref{diciasette}) from which we find that the solution $\Psi_I$ given by (\ref{tredici}) will be valid provided that $\widetilde{a}\gg 10^{-75}$ and $|\widetilde{a}-\widetilde{a}_{-}|\gg 10^{-75}$. Similarly, it can be checked that $\psi_{II}$ and $\Psi_{III}$ holds whenever $|\widetilde{a}-\widetilde{a}_{-}|\gg 10^{-75}$ and $|\widetilde{a}-\widetilde{a}_{+}|\gg 10^{-77}$.
Let us now briefly explore a global upper bound for the tunneling probability. We will begin by noting that as $\widetilde{a}$ approaches infinity, the wave function does not vanish asymptotically, as it can be evinced from the asymptotic behavior described by (\ref{apr1}). This observation signals that in the present case the conclusion stated in \cite{Frid}, according to which the wave function should vanish within the barrier, cannot be applied. Consequently, we can calculate the tunneling probability using the following formula
\begin{equation}\label{TWKB}
T_{WKB}=\mbox{exp}\left(-\frac{2}{\mu}\int_{\widetilde{a}_{-}}^{\widetilde{a}_{+}}\sqrt{V(\widetilde{a})}d\widetilde{a}\right).
\end{equation}
If we observe that the function $\sqrt{V(\widetilde{a})}$ shares its maximum location, denoted as $\widetilde{a}_M$, with $V(\widetilde{a})$ within the interval $[\widetilde{a}_{-},\widetilde{a}_+]$, we can establish an upper bound for it by using $\sqrt{V(\widetilde{a}_M)}$. Consequently, the integral in (\ref{TWKB}) can be bounded as follows
\begin{equation}
\int_{\widetilde{a}_{-}}^{\widetilde{a}_{+}}\sqrt{V(\widetilde{a})}d\widetilde{a}\leq\sqrt{V(\widetilde{a}_M)}(\widetilde{a}_+-\widetilde{a}_{-}).
\end{equation}
At this point, a comment is in order. Firstly, the right-hand side in the above expression depends on the parameter $\epsilon$. Secondly, as indicated in Table~\ref{tableEins}, the values of $\epsilon$ associated with astrophysically relevant black holes are tiny. This observation leads us to the reasonable conclusion that it is justifiable to expand the combination $\sqrt{V(\widetilde{a}_M)}(\widetilde{a}_+-\widetilde{a}_{-})$ with respect to the small parameter $\epsilon$.
\begin{table}[ht]
\caption{Numerical values of the parameter $\epsilon$ defined in (\ref{pp}) in the case of different black hole scenarios. Here, $M_\odot=1.989\cdot 10^{30}$ Kg denotes the solar mass.}
\begin{center}
\begin{tabular}{ | l | l | l | l|l|l|}
\hline
$\mbox{Black Hole}$ & $M/M_\odot$  & $\epsilon$    \\ \hline
\mbox{TON618}       & $6.6\cdot 10^{10}$ & $2.8\cdot 10^{-24}$\\ \hline
\mbox{Sagittarius A$^{*}$} & $4.1\cdot 10^{6}$  & $10^{-32}$\\ \hline
\mbox{GW170817}     & $2.74$  & $4.8\cdot 10^{-45}$\\ \hline
\end{tabular}
\label{tableEins}
\end{center}
\end{table}
More precisely, by means of (\ref{apm}), (\ref{locaM}) and (\ref{VaM}) we find that
\begin{equation}
 \sqrt{V(\widetilde{a}_M)}(\widetilde{a}_+-\widetilde{a}_{-})=\frac{1}{2\epsilon}\left[1-\left(\frac{3}{2}+\sqrt{2}\right)\sqrt{\epsilon}+\mathcal{O}(\epsilon)\right].
\end{equation}
Given the very small magnitude of $\epsilon$, we can bound the tunneling probability  from above by
\begin{equation}
    T_{WKB}\approx
    \mbox{exp}\left(-\frac{\sqrt{\beta}}{\epsilon}\right)=\mbox{exp}\left(-\frac{9\rho_{Pl}}{8\rho_{vac}}\right)\sim\mbox{exp}\left(-10^{123}\right).
  \end{equation}
We acknowledge here the extremely small value of the tunneling
  probability. We attribute it to the fact that the tunneling seems to
  be macroscopic.  Indeed, a positive cosmological constant causes the
  effective potential of the radial geodesic equation of motion to
  develop a local maximum far away from the center
  \cite{deSitter}. The same happens in the WDW equation.
 
The second notable feature of this upper bound is that it does not show any dependence on the physical parameters of the object which underwent the gravitational collapse. Moreover, in the realm of black hole to white hole transitions, various models and theories have been proposed to understand the underlying mechanisms. The present approach, which rests on the application of the WDW equation in the context of gravitational collapse in a universe with a positive cosmological constant, predicts a very small tunneling probability for a black hole to emerge as a white hole, on the order of  $\mbox{exp}(-10^{123})$. 

\subsection{Dwell time or residence time in the barrier}
This section has a twofold purpose.  In spite of the fact that
  the WDW equation has no time, we would like to show that it is
  possible to introduce non-classical time concepts based on quantum
  mechanics. Secondly, the estimate of the so-called tunneling time
  will reinforce our conclusion about the macroscopic nature of the tunneling.
  
Apart from the more commonly evaluated barrier penetration or the 
transmission probability discussed above, 
there exists a lot of interest in the time spent by the tunneling 
particle in the allowed and forbidden regions in tunneling. 
Indeed, starting with the pioneering work of Wigner on the 
phase time delay \cite{wigner}, there exist several definitions such as 
the Larmor, traversal, residence or dwell time, and even complex 
times, which are often derived  
within the framework of semi-classical approximations (see 
\cite{hauge} for a review of these concepts). Among the various definitions, 
the dwell time seems to have found a straightforward interpretation in 
terms of physically measurable quantities such as the lifetimes of decaying 
nuclei \cite{Kelkar2007,NMEPL2009}.    
In the context of the present work, an estimate of the dwell time in the 
barrier region may allow us to comment on the possibility of the transition 
from the black hole to a white hole.

In order to briefly introduce the concept, 
let us consider an arbitrary barrier $V(x)$ in one-dimension 
confined to an interval 
$(x_1, x_2)$. The dwell time in the barrier is given as
\begin{equation}\label{dwellt} 
\tau_D\,=\frac{\int_{x_1}^{x_2}\,|\Psi(x)|^2\,dx}{j}\, .
\end{equation}
Here $\Psi(x)$ is the time independent solution of the Schr\"odinger equation 
in the given region and $j$ the current density. 
In what follows, we shall evaluate the dwell
time in a semiclassical 
approximation for the wave function in the present work. However, 
one can always 
define a dwell time as above in a given region of space, be it with an exact 
or an approximate wave function. 
Considering the analogy between the standard time 
independent Schr\"odinger equation in one-dimension 
and equation (\ref{pora}), namely,
\begin{equation}\label{pora2}
-\frac{\mu^2}{L_0^2} \frac{d^2\Psi}{da^2}+V(a/L_0)\Psi(a/L_0)=0\,,
\end{equation}
we can write the ``average dwell time" in the forbidden 
region (under the barrier) as \cite{NMEPL2009}
\begin{equation}\label{dwellt2}
\tau_D^{\rm II} \, = \, 
\frac{L_0}{2 \mu}  
\, \int_{\tilde{a}_{-}}^{\tilde{a}_{+}} \, 
\frac{d\tilde{a}}{\sqrt{V(\tilde{a})}} \, {\rm exp} \,\biggl 
[ - 2 \, \int_{\tilde{a}_{-}}^{\tilde{a}} \, 
da^{\prime} \, \sqrt{V(a^{\prime})}\biggr ] \, , 
\end{equation}  
where we have substituted for $\tilde{a} = a/L_0$ in (\ref{pora}). 
Note that the normalization of the wave function cancels with the same factor
appearing in the expression for the current density, and hence, the dwell time 
can be evaluated simply with the knowledge of the potential barrier and the 
classical turning points. 

In order to get an estimate of the order of magnitude of the dwell 
time in the barrier region, let us consider the barrier to be rectangular, with 
the height given by its maximum value. We would therefore obtain a lower limit 
on the dwell time with the substitution, $V(\tilde{a}) \to V(\tilde{a}_M)$. 
Thus, the lower bound on the dwell time in region II is given by
\begin{equation}
\tau_D^{\rm II} \, = \,\frac{L_0}{4 \mu V(\tilde{a}_M)}\,\biggl ( 
1 - e^{{-2\sqrt{V(\tilde{a}_M)} (\tilde{a}_+ - \tilde{a}_{-})}} \biggr )\,.
\end{equation}
With $L_0 = 1/\sqrt{k} = 3/(8 \pi G \rho_0) = 4.23 \times 10^{12}$ s and 
using the values of the barrier from Figure \ref{figure1}, 
$\tau_D^{\rm II}$ has a lower bound of 
approximately $10^{113}$ seconds or $10^{96}$ billion years.

Equation (\ref{dwellt}) is, in principle, the definition of an average dwell time or
the average amount of time spent in the region 
$(x_1, x_2)$ regardless of how the particle escaped (by reflection or 
transmission). 
The transmission and reflection dwell times for the particular cases when the 
particle is bound in the barrier and later either got transmitted or reflected
is written by replacing the current density $j$ by a transmitted and reflected 
flux,  
$j_T = j\, T$ and $j_R = j\, R$ respectively. Here, $T$ and $R$ are the 
transmission and reflection coefficients and $R + T = 1$ due to conservation
of probability.  
One would then obtain \cite{mario}
\begin{equation}
\frac{1}{\tau_D} \, =\, \frac{T}{\tau_D} \, + \, \frac{R}{\tau_D} \, 
= \, \frac{1}{\tau_{D,T}} \, + \, \frac{1}{\tau_{D,R}}.
\end{equation}
The lower bound on $\tau_{D,T}^{II}$ would thus be extremely large.  

This result stands in stark contrast to the models discussed in \cite{Malafarina}. For instance, the spherical shells model mentioned in \cite{116} and further analyzed in \cite{31,33} suggests that the exterior region of a black hole may undergo a transition to a white hole, with the timescales of this transition being relatively short. Furthermore, the outer horizon undergoes three distinct stages before matter starts to emerge again: the stages of being a black hole, transitioning, and then becoming a white hole. The phase where it acts as a black hole persists until the shift towards the white hole phase starts. This shift can happen almost instantly or take a longer duration, leading to an intermediate state termed by \cite{116} as the 'gray horizon'—a blend of the black and white hole phases. Once this transition ends, the system stabilizes into a white hole state. To be precise, the duration for which the horizon remains a black hole is the period during which distant observers perceive it as such. However, it is worth mentioning that in models where time is symmetrical, a black hole horizon that exists for an extended period suggests a similarly prolonged existence for the white hole horizon. This prolonged white hole phase could be problematic because of recognized instabilities in white hole configurations, as pointed out by \cite{124,125}. 

The authors in \cite{32} constructed a metric representing the black hole tunneling  into a white hole. This was achieved by employing classical equations outside the quantum region, estimating the onset of quantum gravitational phenomena, and considering indirect evidence of quantum gravity's effects. It is important to note, as highlighted by the authors, that this approach is not derived from first principles. A comprehensive theory of quantum gravity would be required for such a derivation. One of the key aspects of this model is that the time it takes for this transition to occur is influenced by the Schwarzschild mass of the black hole, more precisely, the larger the Schwarzschild mass, the longer it might take for the black hole to transition into a white hole. On the other hand, if we use a theory of quantum gravity such as the one represented by the WDW equation, and we compare the model in \cite{32} with the predictions emerging from our approach, there is a stark contrast. The WDW equation suggests that the likelihood of a black hole transitioning into a white hole is exceedingly rare, almost to the point of being negligible. 

In \cite{Batta}, the study focused on quantum tunneling from black holes, exploring the theoretical possibility of particles tunneling into their time-reversed counterparts, i.e. white holes. The dynamics of black-to-white hole bounces from matter collapse was further analyzed by \cite{Anc}.

We conclude this section by mentioning that \cite{31,33,99,122} introduced the concept of 'oscillating' black holes, which periodically transition between black and white hole states. Such periodic behavior, if validated, could introduce novel astrophysical phenomena and challenge our current understanding of black holes. However, the WDW equation, when applied to gravitational collapse, casts doubt on these oscillating solutions because it suggests that the probability of black to white hole transition is extremely low. This indicates that the likelihood of observing oscillating black holes might be minimal. On the other hand, in light of the implications emerging from the WDW equation, the idea of oscillating black holes, while intriguing, requires further rigorous investigation and empirical validation.

\section{Conclusions}
The Friedmann-Robertson-Walker metric is applicable to expanding or
contracting universes as well as to collapsing matter. This fact makes
the application of the WDW equation in the late
gravitational collapse an appealing tool to study quantum effects
in a black hole after all matter has entered the horizon. Indeed, many
features of the WDW equation in collapse resemble the application of
WDW in cosmology. In particular, the equation becomes a timeless
Schr\"{o}dinger-like equation of the form $H\Psi=0$. Assuring the
correct Lagrangian with matter suitable for quantization, we have taken up the task to examine the above-mentioned quantum effects, paying attention to the ambiguity
of  factor ordering. In the simplest case where the kinetic term
becomes a double derivative with respect to the scale factor,
the solution of the wave function is related to the one dimensional
harmonic oscillator. Although the WDW equation is not inherently an eigenvalue problem, it results in a quantization of the density in terms of the Planck density. We reach the classical limit in the case of a large quantum number $n$. The probability density displays
certain preferred values, which makes the spacetime look quantized in
probabilistic sense. The central singularity is avoided. The question
arises whether these nice properties change if we introduce more
complicated factor orderings. We solve the corresponding WDW equations
just by the standard ansatz of an exponential multiplied by
polynomials. We can construct these polynomials explicitly and at the
same time obtain a quantization condition which is similar to the
original one in the simplest case.  The probability distribution has
again preferred values, even though the latter are distributed in a
different manner as compared to the simplest case. The classical central
singularity does not appear.

A tunneling scenario emerges when we switch on a positive
cosmological constant. The barrier which gives rise to the tunneling
is also encountered in classical gravity \cite{deSitter}. We could
give an estimate of the tunneling probability, which turns out to be
very small. This is not surprising, as we talk here about a macroscopic
tunneling.

It is satisfying to see these features coming out of the WDW equation
applied to collapse. It might take a long time till we settle down the
full theory of quantum gravity, and even longer till we can confirm it
by observations. But we have certain expectations what such a theory
would predict.  For instance, we would indeed expect the central
singularity is absent in the quantum version.  Secondly, we might also
suspect that the spacetime will appear if not fully discretized, at
least taking probabilistically preferred values. Thirdly, as in the
standard quantum mechanics, one would expect some quantities to be
quantized.

Last but not least, we have focused on the quantum gravitational effects during the collapse process, as described by the WDW equation. While Hawking radiation is a significant effect, its impact on the immediate dynamics of collapse is negligible within the WDW formalism. A comprehensive treatment of Hawking radiation would require a different approach beyond the scope of this work.

\begin{acknowledgments}
N.G.K. thanks the Faculty of Science, Universidad de Los Andes,
Colombia, for financial support through Grant No. INV-2023-162-2841.
We would like to thank Prof. R. Baier for useful discussions on the
self-adjointness of the WDW operator.
\end{acknowledgments}


\begin{thebibliography}{99}
\bibitem{Weinberg}
S. Weinberg, {\it Gravitation and Cosmology: Principles and Applications of The General Theory of Relativity}, John Wiley \& Sons (1972).

\bibitem{Joshi}
P. S. Joshi, {\it Gravitational Collapse and Spacetime Singularities}, Cambridge University Press (2012).

\bibitem{Kent}
B. K. Harrison, K. S. Thorne, M. Wakano and J. A. Wheeler, {\it Gravitation Theory and Gravitational Collapse}, Univ of Chicago Press (1965).

\bibitem{compactobjects}
S. L. Shapiro and S. A. Teukolsky, {\it Black Holes, White Dwarfs and Neutron Stars: The Physics of Compact Objects}, John Wiley \& Sons (1983).

\bibitem{Hawking}
S. Hawking,  {\it{Black hole explosions?}}, Nature {\bf{248}}, 30 (1974); 
S. Hawking,  {\it{Particle creation by black holes}}, Commun. Math. Phys. {\bf{43}}, 199 (1975).

\bibitem{Bambi}
C. Bambi, {\it{Regular Black Holes: Towards a New Paradigm of Gravitational Collapse}},  Springer Series in Astrophysics and Cosmology (2023); 
C. Bambi, {\it{Black Holes: A Laboratory for Testing Strong Gravity}}, Springer Nature (2017).

\bibitem{Malafarina}
D. Malafarina, {\it{Black Hole Bounces on the Road to Quantum Gravity}}, Universe {\bf{4}}, 92 (2018); 
D. Malafarina, {\it{Classical collapse to black holes and quantum bounces: A review}}, Universe {\bf{3}}, 48 (2017); 
H. Chakrabarty, A. Abdujabbarov, D. Malafarina and C. Bambi, {\it{A toy model for a baby universe inside a black hole}}, Eur. Phys. J. C {\bf{80}}, 373 (2020); 
D. Malafarina, {\it{A Brief Review of Relativistic Gravitational Collapse}}, Astrophys. Space Sci. Libr. {\bf{440}}, 169 (2016).

\bibitem{QuantumGravity}
C. Kiefer, {\it{Quantum Gravity}}, Oxford University Press, Oxford, 2nd edition (2012); 
C. Rovelli, {\it{Quantum Gravity}}, Cambridge University Press, Cambridge (2004).

\bibitem{Rocci}
A. Rocci, {\it{On first attempts to reconcile quantum principles with gravity}}, J. Phys. Conf. Ser. {\bf 470}, 012004 (2013): 
G. Peruzzi and A. Rocci, {\it{Tales from the prehistory of Quantum Gravity}}, Eur. Phys. J. H {\bf 43}, 185 (2018).

\bibitem{Rovelli}
C. Rovelli, {\it{Notes for a brief history of quantum gravity}} in Proceedings of the MGIX MM Meeting, The University of Rome ‘La Sapienza’,  edited by V. G. Gurzadyan, R. T. Jantzen and R. Ruffini, World Scientific, Singapore (2002), arXiv:gr-qc/0006061 (2000).

\bibitem{Esposito}
B. C. De Witt and G. Esposito, {\it{An introduction to quantum gravity}}, Int. J. Geom. Meth. Mod. Phys.{\bf 5}, 101 (2008).

\bibitem{ADM}
T. Thiemann,  {\it{Canonical Quantum Gravity, Constructive QFT and Renormalisation}}, Front. Phys. {\bf 8},  457 (2020); F. Cianfrani, O. M. Lecian, M. Lulli and G. Montani, {\it{Canonical Quantum Gravity}}, World Scientific, Singapore (2014).

\bibitem{WDW}
B. S. DeWitt, {\it{Quantum Theory of Gravity. I. The Canonical Theory}}, Phys. Rev. {\bf{160}}, 1113 (1967); J. A. Wheeler, in Batelle Rencontres: {\it{1967 Lectures in Mathematics and Physics}}, edited by C. DeWitt and J. A. Wheeler (Benjamin, New York, 1968), p. 242.

\bibitem{WDWpapers1}
T. Padmanabhan and T. P. Singh, {\it{On the semiclassical limit of the Wheeler-DeWitt equation}}, Class. Quantum Grav. {\bf 7} 411 (1990);
E. Adi and S. Soloma, {\it{The solution to Wheeler-de Witt equation is eight}}, Phys. Lett. {\bf 336}, 152 (1994); 
E. Rodrigo, {\it{Solving the Wheeler- de Witt equation for Kaluza-Klein theories}}, Phys. Lett. {\bf B160}, 43 (1985); 
J. W. Norbury, {\it{From Newton's law to Wheeler-de Witt equation}}, Eur. J. Phys. {\bf 19}, 43 (1998); 
G. Giampieri, {\it{A new solution of the Wheeler- de Witt equation}}, Phys. Lett. {\bf B261}, 411 (1991); 
J. Kowalski-Glikman and K. A. Meissner, {\it{A class of exact solutions of the Wheeler-de Witt
equation}}, Phys. Lett. {\bf B376}, 48 (1996).

\bibitem{WDWpapers2}
R. Garattini, {\it{Extracting Maxwell charge form Wheeler-de Witt equation}}, Phys. Lett. {\bf B666}, 189 (2008); 
W. Nelson and M. Sakellariadan, {\it{On the possibility of Dark Energy from corrections to the Wheeler-De Witt equation}}, Phys. Lett. B {\bf 661}, 37 (2008); 
M. Pav\v si\v c, {\it{Klein-Gordon-Wheeler-de Witt-Sch\"odinger Equation}}, Phys. Lett. B {\bf 703}, 614 (2011); 
M. Pav\v si\v c, {\it{Wheeler-DeWitt Equation in Five Dimensions and Modified QED}}, Phys. Lett. B {\bf 717}, 441 (2012); 
T. Kubota, T. Ueno and N. Yokoi, {\it{Wheeler-DeWitt Equation in AdS/CFT Correspondence}}, Phys. Lett. B {\bf{579}}, 200 (2004).

\bibitem{WDWpapers3}
C. Rovelli, {\it{The strange equation of quantum gravity}}, Class. Quantum Grav. {\bf{32}}, 124005 (2015); 
C. Chowdhury, V. Godet, O. Papadoulaki and S. Raju, {\it{Holography from the Wheeler-DeWitt equation}}, J. High Energ. Phys. {\bf{2022}}, 19 (2022); 
Ch. F. Steinwachs and M. L. van der Wild, {\it{Quantum gravitational corrections from the Wheeler-DeWitt equation for scalar-tensor theories}}, Class. Quantum Grav. {\bf{35}}, 135010 (2018); 
P. T. Shestakova, {\it{Is the Wheeler–DeWitt equation more fundamental than the Schr\"{o}dinger
equation?}}, Int. J. Mod. Phys. D {\bf 27}, 1841004 (2018);
R. Baier and C. Peterson, {\it{Massive and Massless Quantum Cosmos}}, arXiv:gr-qc/2312.17647 (2023).

\bibitem{Chen1}
M. Bouhmadi-Lopez, S. Brahma, C.-Y. Chen, P. Chen and D.-H. Yeon, {\it{Annihiation-to-nothing: a quantum gravity boundary condition for the Schwarzschid blakc hole}}, JCAP {\bf 11}, 002 (2020).
 
\bibitem{Chen2}
C.-H. Chien, G. Tumurtushaa and D.-H. Yeon, {\it{Wheeler-De Witt equation beynd the cosmological horizon: Annihilation to nothing avoidance, and loss of quantum coherence}}, Phys. Rev. D {\bf 108}, 10023539 (2023).

\bibitem{WDWcosmology}
V. Vilenkin, {\it{Approaches to Quantum Cosmology}}, Phys. Rev. D {\bf{50}}, 2581 (1994);
M. Znojil, {\it{Wheeler-DeWitt Equation and the Applicability of Crypto-Hermitian Interaction Representation in Quantum Cosmology}}, Universe {\bf{8}}, 385 (2022); 
J. W. Norburg, {\it{Some connection between Quantum tunnelling and Inflation}},  Phys. Lett. B {\bf 433}, 223
(1998); 
W. Fischler, B. Ratra and L. Susskind, {\it{Quantum Mechanics in Inflation}}, Nucl. Phys. B {\bf 259}, 730 (1983);  
D. He and Q.-Y. Cai, {\it{Wheeler- De Witt equation rejects quantum effects of grown-up universes as a candidate for dark energy}}, Phys. Lett. B {\bf 809}, 135747 (2020);
D.L. Wiltshire, {\it{An introduction to quantum cosmology}}, arXiv:gr-qc/0101003 (2000).
\bibitem{Mathur1}
S. D. Mathur, {\it{The fuzzball proposal for black holes: an elementary review}}, Fortsch. Phys. {\bf{53}}, 793 (2005).
\bibitem{Mathur2}
S. D. Mathur, {\it{Fuzzballs and the information paradox: A summary and conjectures}}, Adv. Sci. Lett. {\bf{2}}, 133 (2008).
\bibitem{Bena}
I. Bena and N. P. Warner, {\it{Black Holes, Black Rings and their Microstates}}, Lect. Notes Phys. {\bf{755}}, 1 (2008)
\bibitem{Struyve2017}
W. Struyve, {\it{Loop quantum cosmology and singularities}}, Sci. Rep. {\bf{7}}, 8161 (2017).
\bibitem{Bojowald2007}
M. Bojowald, {\it{Singularities and quantum gravity}}, lecture course at the XIIth Brazilian School on Cosmology and Gravitation, September 2006, https://arxiv.org/abs/gr-qc/0702144
\bibitem{Singh2024}
H. Singh and M. K. Nandy, {\it{Black hole singularity resolution in Wheeler-DeWitt quantum gravity}}, Ann. Phys. {\bf{468}}, 169719 (2024).
\bibitem{ourpaper}
D. Batic and M. Nowakowski, {\it{Gravitational collapse via the Wheeler-de Witt equation}}, Ann. Phys. {\bf 461}, 169574 (2024).

\bibitem{Rosen}
N. Rosen, {\it{Quantum mechanics of a miniuniverse}},  Int. J. Theor. Phys. {\bf{32}}, 1435 (1993).

\bibitem{Feleppa}
C. Corda and F. Feleppa, {\it{Quantum black hole as a gravitational atom}}, Adv. Theor. Math. Phys. {\bf{10}}, 3537 (2022). 

\bibitem{time1}
C. J. Isham, {\it{Canonical Quantum Gravity and the Problem of Time}}, in {\it{Integrable Systems, Quantum Groups, and Quantum Field Theories}}, Eds. L. A. Ibort and M. A. Rodriguez, Springer Verlag (1993).

\bibitem{time2}
K. V. Kuchar, {\it{Time and interpretations of quantum gravity}}, in Proceedings of the 4th Canadian Conference on General Relativity and Relativistic Astrophysics, Eds. G. Kunstatter, D. Vincent, and
J. Williams, World Scientific Publishing Company, Singapore (1992).

\bibitem{time3}
C. Kiefer and P. Peter {\it Time in Quantum Cosmology}, Universe 2022, {\bf 8}, 36  (2022).

\bibitem{time4}
J. B. Barbour, {\it{The timelessness of quantum gravity: I. The evidence from the classical theory}}, Class. Quantum Gravity {\bf 11}, 2853 (1994); 
J. B. Barbour, {\it{The timelessness of quantum gravity: II. The appearance of dynamics in static configurations}}, Class. Quantum Gravity {\bf 11}, 2875 (1994);
J. B. Barbour, {\it{The end of time: the next revolution in physics}}, Oxford University Press, Oxford (1999).

\bibitem{time5} 
J. B. Barbour, B. Z. Foster and N. O. Murchadha, {\it{Relativity without relativity}},  Class. Quantum Gravity {\bf 19}, 3217 (2002).

\bibitem{time6} 
E. Y. Chua and C. Callender, {\it{No time for time from no-time}}, Philos. Sci. {\bf 88}, 1172 (2021).

\bibitem{time7}
S. B. Gryb, {\it{Jacobi's Principle and the Disappearance of Time}}, Phys. Rev. D {\bf 81}, 044035 (2010).

\bibitem{time8}
C. Rovelli, {\it{Quantum mechanics without time: A model}}, Phys. Rev. D {\bf{42}}, (1990), 2638; 
C. Rovelli, {\it{Time in quantum gravity: An hypothesis}},  Phys. Rev. D {\bf{43}}, (1991), 442.

\bibitem{time9}
A. M. Frauca, {\it{Reassessing the problem of time of quantum gravity}}, Gen. Relativ. Gravit. {\bf 55}, 21 (2023).

\bibitem{time10}
D. N. Page and W. K. Wooters, {\it{Evolution without time: Dynamics described by stationary variables}},  Phys. Rev. D {\bf{27}}, 2885 (1983).

\bibitem{time11}
A. Vilenkin, {\it{Boundary conditions in quantum cosmology}}, Phys. Rev. D {\bf{33}}, 3560 (1986).
C. Kiefer and P.Peter, {\it{Time in quantum cosmology}}, Universe 2022, {\bf 8}, 36 (2022).

\bibitem{time12}
L. Smolin, {\it{Time Reborn: From the Crisis in Physics to the Future of the Universe}},  Mariner Books  (2014); 
E. Anderson, {\it The Problem of Time: Quantum Mechanics Versus General Relativity}, Springer Verlag (2017).
 
\bibitem{Stojkovic1}
E. Greenwood and D. Stojkovic, {\it{Quantum gravitational collapse: Non-singularity and non-locality}}, JHEP {\bf{06}}, 042 (2007).

\bibitem{Stojkovic2}
T. Vachaspati and D. Stojkovic, {\it{Quantum radiation from quantum gravitational collapse}},  Phys. Lett. B  {\bf 663}, 107 (2008).

\bibitem{Pal}
S. Chowdhury, K. Pal, K. Pal and T. Sarkar, {\it{Quantum potential in bouncing dust collapse with a negative cosmological constant}}, Phys. Lett B {\bf 816}, 136269 (2021).

\bibitem{Bode}
P. Bodenheimer, {\it{Stellar Structure and Evolution}}, Encyclopedia of Physical Science and Technology (Third Edition), Academic Press (2003).

\bibitem{Schiff}
L. I. Schiff, {\it{Quantum Mechanics}}, McGraw Hill (1968).

\bibitem{KT}
E. W. Kolb and M. S. Turner, {\it{The Early Universe}}, Addison-Wesley (1989).

\bibitem{Inverno}
R. D'Inverno, {\it{Introducing Einstein's Relativity}}, Clarendon Press, Oxford (1998).

\bibitem{Schutz}
B. F. Schutz, {\it{Perfect Fluids in General Relativity: Velocity Potentials and 
a Variational Principle}}, Phys. Rev. D {\bf{2}}, 2762 (1970).

\bibitem{Ray}
J. R. Ray, {\it{Lagrangian Density for Perfect Fluids in General Relativity}}, J. Math. Phys. {\bf{13}}, 1451 (1972).

\bibitem{Braz}
H. S. Vieira and V. B. Bezerra, {\it{Class of solutions of the Wheeler-DeWitt equation in the Friedmann-Robertson-Walker universe}}, Phys. Rev. D {\bf{94}} 023511  (2016).

\bibitem{Grif}
D. J. Griffiths, {\it{Introduction to Quantum Mechanics}}, 2nd edition, Pearson Prentice Hall (2005).

\bibitem{Hartle}
J. B. Hartle and S. W. Hawking, {\it{Wave Function of the Universe}}, Phys. Rev. D {\bf{28}}, 2960 (1983).

\bibitem{Vilenkin}
A. Vilenkin, {\it{Approaches to quantum cosmology }}, Phys. Rev. D {\bf{50}}, 2581 (1994).

\bibitem{Page}
S. W. Hawking and D. N. Page, {\it{Operator ordering and the flatness of the universe}}, Nucl. Phys. B {\bf{264}}, 185 (1986).

\bibitem{Kont}
N. Kontoleon and D.L. Wiltshire, {\it{Operator ordering and consistency of the wave function of the universe}}, Phys. Rev. D {\bf{59}}, 063513 (1999).

\bibitem{Vil}
A. Vilenkin, {\it{Quantum cosmology and the initial state of the Universe}}, Phys. Rev. D {\bf{37}}, 888 (1988).

\bibitem{Gao}
D. He, D. Gao and Q.-Y. Cai, {\it{Dynamical interpretation of the wavefunction of the universe}}, Phys. Lett. B {\bf{748}}, 361 (2015).

\bibitem{Vieira}
H.S. Vieira, V.B. Bezerra, C.R. Muniz, M.S. Cunha and H.R. Christiansen, {\it{Some exact results on quantum relativistic cosmology: dynamical interpretation and tunneling phase}},  Phys. Lett. B {\bf{809}}, 135712 (2020).

\bibitem{He1}
D. He, D. Gao and Q.-Y. Cai, {\it{Spontaneous creation of the universe from nothing}}, Phys. Rev. D {\bf{89}}, 083510 (2014).

\bibitem{He2}
D. He and Q.-Y. Cai, {\it{Inflation of small true vacuum bubble by quantization of Einstein–Hilbert action}}, Sci. China Phys. Mech. Astron. {\bf{58}}, 079801 (2015).

\bibitem{Steigl}
R. \v Steigl and F. Hinterleitner, {\it{Factor ordering in standard quantum cosmology}},  Class. Quantum Gravity {\bf{23}}, 3879 (2006).

\bibitem{deSitter}
A. Balaguera-Antolinez, C. G. Boehmer and M. Nowakowski, {\it{Scales Set by the Cosmological Constant}},  Class. Quantum Gravity {\bf{23}}, 485 (2006).

\bibitem{Bron}
I. N. Bronshtein, K.A. Semendyayev, G. Musiol and H. M\"{u}hlig, {\it{Handbook of Mathematics}}, 6th edition, Springer Verlag Berlin Heidelberg (2015).

\bibitem{Gil} G. Paz, {\it{The non-self-adjointness of the radial
      momentum operator in n dimensions}},  J. Phys. A:
  Math. Gen. {\bf 35} (2002) 3727.


\bibitem{Berezin}
F. A. Berezin and M. A. Shubin, {\it{The Schr\"{o}dinger equation}}, Kluwer Academic Publishers (1991).

\bibitem{McLeod}
J. B. McLeod, {\it{The Limit-Point Classification of Differential Expressions}}, North-Holland Mathematical Studies {\bf{13}}, 57 (1974).

\bibitem{seven}
L. Randall and R. Sundrum, {\it{An Alternative to Compactification}}, Phys. Rev. Lett. {\bf{83}}, 4690 (1999).

\bibitem{four}
S. Kar and R. Parwani, {\it{Can degenerate bound states occur in one-dimensional quantum mechanics?}}, Europhys. Lett. {\bf{80}}, 30004 (2007).

\bibitem{Bender}
C. M. Bender and S. Boettcher, {\it{Real Spectra in Non-Hermitian Hamiltonians Having PT Symmetry}}, PRL {\bf{80}}, 5243 (1998).

\bibitem{Bender1}
C. Bender, {\it{Making sense of non-Hermitian Hamiltonians}}, Rep. Prog. Phys. {\bf{70}}, 947 (2007).

\bibitem{Vil1986}
A. Vilenkin, {\it{Boundary conditions in quantum cosmology}}, Phys. Rev. D {\bf{33}}, 3560 (1986).

\bibitem{Frid}
H. Friedrich and J. Trost, {\it{Working WKB waves far from the semiclassical limit}}, Phys. Rep. {\bf{397}}, 359 (2004).

\bibitem{116}
M. Ambrus and P. Hajicek, {\it{Quantum superposition principle and gravitational collapse: Scattering times for spherical shells}} Phys. Rev. D {\bf{72}}, 064025 (2005).
\bibitem{wigner}
E. P. Wigner, {\it 
Lower Limit for the Energy Derivative of the Scattering Phase Shift}, 
Phys. Rev. {\bf 98}, 145 (1955). 
\bibitem{hauge}
E. H. Hauge and J. A. St\o{}vneng, {\it Tunneling times: A critical review}, 
Rev. Mod. Phys. 61, 917 (1989). 
\bibitem{Kelkar2007}
N. G. Kelkar, {\it Quantum Reflection and Dwell Times of Metastable States}, 
Phys. Rev. Lett. {\bf 99}, 210403 (2007).
\bibitem{NMEPL2009}
N. G. Kelkar, H. M. Casta\~neda and M. Nowakowski, {\it Quantum time scales 
in alpha tunneling}, Eur. Phys. Lett. {\bf 85}, 20006 (2009). 
\bibitem{mario}
M. Goto, H. Iwamoto, V. M. de Aquino, V. C. Aguilera-Navarro and  
D. H. Kobe, {\it Relationship between dwell, transmission and
reflection tunnelling times}, J. Phys. {\bf A 37}, 3599 (2004). 
\bibitem{31}
C. Barceló, R. Carballo-Rubio and L. J. Garay, {\it{Mutiny at the white-hole district}},  Int. J. Mod. Phys. D {\bf{23}}, 1442022 (2014).

\bibitem{33}
C. Barceló, R. Carballo-Rubio, L. J. Garay and G. Jannes, {\it{The lifetime problem of evaporating black holes: Mutiny or resignation}}, Class. Quantum Gravity {\bf{32}}, 035012 (2015).

\bibitem{124}
D. M. Eardley, {\it{Death of white holes in the early universe}}, Phys. Rev. Lett. {\bf{33}}, 442 (1974).

\bibitem{125}
C. Barrabès, P. R. Brady and E. Poisson, {\it{Death of white holes}}, Phys. Rev. D {\bf{47}}, 2383 (1993).

\bibitem{32}
H. M. Haggard and C. Rovelli, {\it{Black hole fireworks: Quantum-gravity effects outside the horizon spark black to white hole tunneling}}, Phys. Rev. D {\bf{92}}, 104020 (2015).

\bibitem{Batta}
B. Chatterjee, A. Ghosh and P. Mitra, {\it{Tunnelling from black holes and tunnelling into white holes}},  Phys. Lett. B . {\bf{661}}, 307 (2008).

\bibitem{Anc}
J. B. Achour, S. Brahma, S. Mukohyama and J.-P. Uzan, {\it{Towards consistent black-to-white hole bounces from matter collapse}}, JCAP {\bf{09}}, 020 (2020).

\bibitem{99}
C. Barcelò, L. J. Garay and G. Jannes, {\it{Quantum Non-Gravity and Stellar Collapse}}, Found. Phys. {\bf{41}}, 1532 (2011).

\bibitem{122}
C. Barceló, R. Carballo-Rubio and L. J. Garay, {\it{Exponential fading to white of black holes in quantum gravity}}, Class. Quantum Gravity {\bf{34}}, 105007 (2017).


\end{thebibliography}
\end{document}